\documentclass{aa}

\usepackage{natbib}
\usepackage{graphicx}
\usepackage{txfonts}
\usepackage[breaklinks=true, colorlinks=blue,linkcolor=black, citecolor=blue]{hyperref}
\usepackage{subfig}
\usepackage{xspace}

\usepackage{euclid}

\setcitestyle{notesep={ }}

\newcommand{\dz}{\ensuremath{\delta_\mathrm{z}}\xspace}
\newcommand{\dg}{\ensuremath{\delta_\mathrm{g}}\xspace}
\newcommand{\dm}{\ensuremath{\delta_\mathrm{m}^{\rm 3D}}\xspace}

\newcommand{\kcmb}{\ensuremath{\kappa_\mathrm{CMB}}\xspace}
\newcommand{\sigmaz}{\ensuremath{\sigma_\mathrm{z}}\xspace}

\newcommand{\lcdm}{\ensuremath{\mathrm{\Lambda CDM}}\xspace}
\newcommand{\wcdm}{\ensuremath{\mathrm{wCDM}}\xspace}

\newcommand{\Dgz}{\ensuremath{\vec{D}_{\rm g,\, z}}\xspace}
\newcommand{\Dg}{\ensuremath{\vec{D}_{\rm g}}\xspace}
\newcommand{\Dz}{\ensuremath{\vec{D}_{\rm z}}\xspace}

\newcommand{\Dgzk}{\ensuremath{\vec{D}_{\rm g,\, z, \, \kcmb}}\xspace}
\newcommand{\Dgk}{\ensuremath{\vec{D}_{\rm g, \, \kcmb}}\xspace}

\graphicspath{{Figures/}}

\begin{document}

    \title{High-resolution tomography for galaxy spectroscopic surveys with angular redshift fluctuations}
    \author{
        L. Legrand \inst{\ref{inst1}}
        \and C. Hern\'andez-Monteagudo \inst{\ref{inst2}}
        \and M. Douspis \inst{\ref{inst1}}
        \and N. Aghanim \inst{\ref{inst1}}
        \and Ra\'{u}l E. Angulo \inst{\ref{inst3}, \ref{inst4}}
    }

    \institute{
        Université Paris-Saclay, CNRS,  Institut d'astrophysique spatiale, 91405, Orsay, France. \\
        \email{louis.legrand@outlook.com} \label{inst1}
        \and Centro de Estudios de F\'isica  del Cosmos de Arag\'on (CEFCA), Unidad Asociada al CSIC, Plaza San Juan, 1, planta 2, E-44001, Teruel, Spain \label{inst2}
        \and Donostia International Physics Centre (DIPC), Paseo Manuel de Lardizabal 4, E-20018 Donostia-San Sebastian, Spain \label{inst3}
        \and IKERBASQUE, Basque Foundation for Science, E-48013, Bilbao, Spain \label{inst4}
    }

\date{\today}


\abstract{In the context of next-generation spectroscopic galaxy surveys, new statistics of the distribution of matter are currently being developed. Among these, we investigated the angular redshift fluctuations (ARF), which probe the information contained in the projected redshift distribution of galaxies.
Relying on the Fisher formalism, we show how ARF will provide complementary cosmological information compared to traditional angular galaxy clustering. We tested both the standard \lcdm model and the \wcdm extension.
We find that the cosmological and galaxy bias parameters express different degeneracies when inferred from ARF or from angular galaxy clustering. As such, combining both observables breaks these degeneracies and greatly decreases the marginalised uncertainties by a factor of at least two on most parameters for the \lcdm and \wcdm models.
We find that the ARF combined with angular galaxy clustering provide a great way to probe dark energy by increasing the figure of merit of the $w_0$-$w_{\rm a}$ parameter set by a factor of more than ten compared to angular galaxy clustering alone.
Finally, we compared ARF to the CMB lensing constraints on the galaxy bias parameters. We show that a joint analysis of ARF and angular galaxy clustering improves constraints by $\sim 40\%$ on galaxy bias compared to a joint analysis of angular galaxy clustering and CMB lensing.}

\keywords{Cosmology: large-scale structure of Universe - observations - cosmological parameters - dark energy}

\maketitle


\section{Introduction}

In the coming years, large-scale optical and infrared (IR) surveys will map our Universe from the present epoch up to when it was roughly one tenth of its current age with unprecedented accuracy. A significant part of these surveys will be spectroscopic; for example, DESI \citep{DESI_Collaboration_2016}, 4MOST \citep{4most}, WEAVE \citep{weave}, and NISP aboard \Euclid \citep{LaureijsAmiaux2011}, and they will provide us with spectra for large samples of sources. Such spectra will not only enable deep insight into the physics of those objects, but it will also yield accurate estimates of their redshift and thus of their distance to the observer. From the cosmological point of view, this will enable a precise (statistical) characterisation of the (apparent) spatial distribution of those luminous tracers (via  two- or three-point statistics), and this itself should shed precious light on open topics such as the nature of dark energy, the possible interplay of dark energy and dark matter, the mass hierarchy of neutrinos, or possible deviations of gravity from general relativity, to name a few.

At the same time, a different family of surveys will scan the sky at greater depths with optical filters and exquisite image quality. These photometric experiments build very large and high-quality source catalogues, with, however, relatively rough redshift estimations given their moderate number of filters. While mining the faint Universe, these types of surveys will be particularly sensitive, from a cosmological perspective, to the angular clustering of luminous matter, the cosmological aspects of gravitational lensing throughout cosmic epochs, the satellite population in halos, and the formation and evolution of the population of galaxy clusters. In this context, the Dark Energy Survey \citep[DES,][]{AbbottAbdalla2018} is currently providing state-of-the-art cosmological constraints in the late universe, and these should be further complemented by the Vera Rubin Observatory \citep[LSST,][]{IvezicKahn2019}, which, at the same time, will also explore the variability of the night sky in a regime of depth and time domain that remains practically unexplored to date.

An intermediate, third class of experiments also exists. These are the spectro-photometric surveys that conduct standard photometry in a relatively large set (from $\sim 10$ up to $\sim 60$) of narrow-band optical filters. This strategy combines the indiscriminate character of the photometric surveys with {\em high} precision redshift estimates ($\Delta z / (1+z) \sim 10^{-3}$--$10^{-2}$) for a large fraction ($> 20$--$30~\%$) of the detected sources. Given its multi-colour character, these surveys are able to provide pseudo- and photo-spectra in each pixel of the surveyed area. The pioneer example of COMBO-17 has been or is being followed by other efforts such as COSMOS \citep{cosmos}, ALHAMBRA \citep{alhambra}, SHARDS \citep{shards}, PAU \citep{pau}, J-PAS \citep{jpas}, SPHEREx \citep{spherex}, and J-PLUS \citep{jplus}.

In this work, we forecast the cosmological constraints for upcoming spectroscopic and spectro-photometric surveys. In these types of surveys, it is customary to convert redshift estimates into radial distances under the assumption of a given fiducial cosmological model. Angular and redshift coordinates are thus converted into 3D space, where standard 3D clustering analysis techniques are applied.

In our case, however, we chose to follow a different strategy. We focused on a new cosmological observable, namely the angular redshift fluctuations \citep[ARF;][hereafter HMCMA]{arf_lett1}. Being a 2D observable, the ARF field can easily be cross-correlated with other 2D observables, such as the 2D galaxy density filed and the CMB lensing fields.
As shown in HMCMA, ARF are sensitive to the variation of matter density and velocity along the line of sight, while galaxy density is sensitive to the the average or monopole of matter density and velocity in the same redshift range.
ARF present other interesting features, such as being correlated to the cosmic, radial, peculiar velocity fields, or being particularly insensitive to additive systematics that remain constant under the redshift shell subject to analysis  \citep[HMCMA; ][]{arfxkSZ_jonas}.

In this work, we applied the Fisher formalism to the angular galaxy clustering, the ARF, and the CMB lensing convergence observables, and we explored their sensitivity to cosmology in two different observational setups, mimicking those expected for the DESI and \Euclid surveys. We considered the CMB lensing convergence field among our observables, since it constitutes an intrinsically different probe, of which the dependence on the parameters defining the galaxy sample is different from that of angular galaxy clustering and ARF.
Our scope is to assess whether the ARF field can provide complementary information on the galaxy density field and on the CMB lensing field.

The paper is organised as follows. We introduce the spectroscopic galaxy surveys and CMB experiments that we used in our analysis in Sect.~\ref{sec:surveys}. In Sect.~\ref{sec:observables}, we present the angular galaxy clustering, the ARF, and the CMB lensing convergence field. In Sect.~\ref{sec:ston}, we compute the foreseen signal-to-noise ratios of these probe combinations, whilst also introducing the covariance among those observables. In Sect.~\ref{sec:fisher}, we present the predicted constraints on cosmological parameters in the fiducial $\Lambda$CDM scenario. Finally, we discuss our findings in Sect. ~\ref{sec:discussion} and conclude in Sect.~\ref{sec:conclusion}.

Throughout this paper, we use the \Planck 2018 cosmology as our fiducial cosmology. We take the values given in Table 2, column 6 (best-fit with BAO) of \citet{Planck2018CosmoParameters}.
We use the following naming conventions: \textit{observable} refers to a spherical 2D field built on measured quantities such as counts, redshifts, or deflection angles, while \textit{probe} refers to the combination of one or two observables in a given set of summary statistics. In practice, our probes will be the two-point angular power spectra $C_\ell$.
The redshift due to the Hubble expansion is denoted by $z$, while $z_{\rm obs}$ is the measured redshift (which includes redshift distortions induced by radial peculiar velocities). $\Omega_{\rm m,0}$ is the density of matter at $z=0$ in units of the critical density, and $H_0$ is the Hubble constant. $r(z) = \int  \diff z \; c \, / \, H(z)$ is the line of sight comoving distance, and $\diff V_{\Omega} =  \diff V / \diff \Omega  =  r^2 \, \diff r = r^2 (z) \, c/H(z) \, \diff z$ is the comoving volume element per solid angle, with $\diff \Omega$ being a differential solid angle element.
Vectors are in bold font, and a hat denotes a unit vector.


\section{Surveys}
\label{sec:surveys}

\begin{table*}[ht!]
    \centering
    \begin{tabular}{lll}
    \hline
    \hline
        Survey  & \Euclid & DESI  \\
    \hline
        Survey area     & 15\,000~deg$^2$  & 14\,000~deg$^2$   \\
        Redshift estimation  & Slitless spectroscopy & Optical fibre spectroscopy  \\
        Targets & $H_\alpha$ emission line & [OII] doublet     \\
        Redshift range  & $0.9<z<1.8$   & $0.6<z<1.6$ \\
        Average number of galaxies  &  1950~gal~deg$^{-2}$ &  1220~gal~deg$^{-2}$    \\
        Galaxy bias & $b_\mathrm{g}(z)  = 0.79 + 0.68 \, z$ &  $b_\mathrm{g}(z) = 0.84/D(z)$   \\
        Reference            & \cite{EuclidCollaborationBlanchard2019} & \cite{DESI_Collaboration_2016}   \\
    \hline
    \end{tabular}
    \caption{Specifications for the two galaxy surveys under consideration.}
    \label{tab:gal_spec}
\end{table*}

\begin{figure*}[th]
    \centering
    \subfloat[DESI]{
    \includegraphics[width=0.5\textwidth]{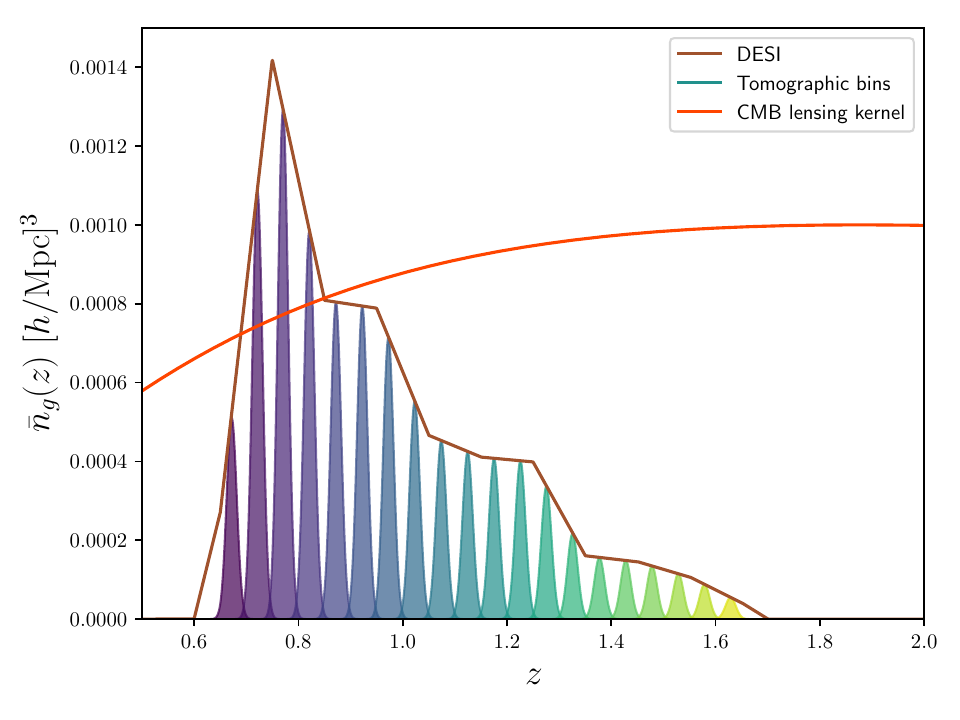}
    \label{fig:nofz_desielg}}
    \subfloat[\Euclid]{
    \includegraphics[width=0.5\textwidth]{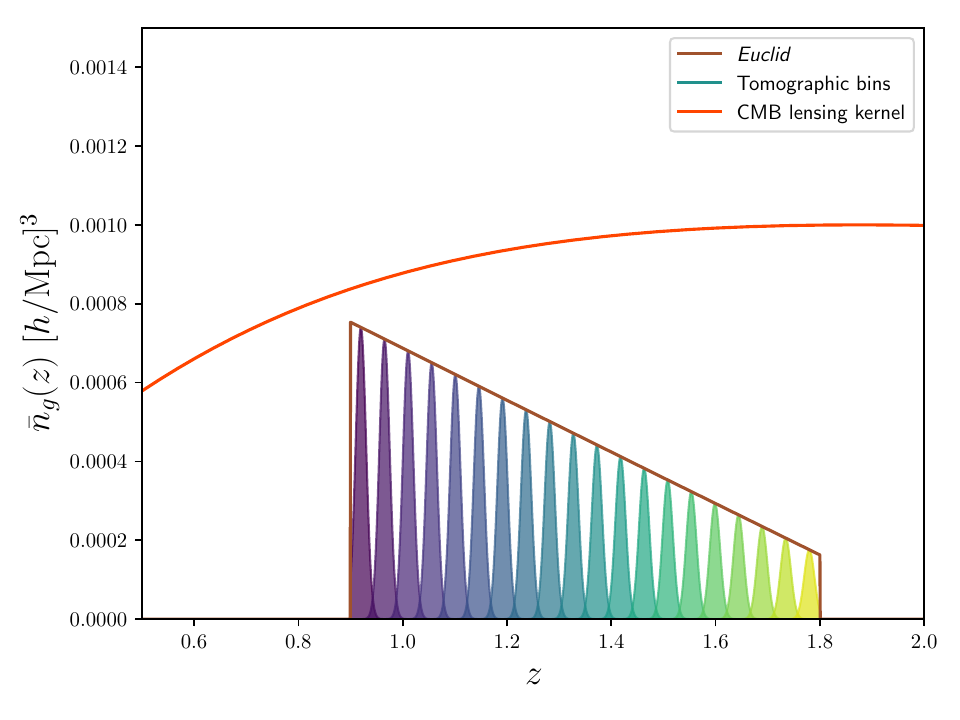}
    \label{fig:nofz_euclid}}
    \caption{Galaxy density distribution as a function of redshift for the emission line galaxies (ELG) of DESI (left panel), and the \Euclid spectroscopic sample (right panel). The filled coloured lines show the Gaussian bins used in our analysis, colour-coded as a function of the bin index. The orange line shows the CMB lensing efficiency kernel (with arbitrary normalisation).}
\end{figure*}

Among the wealth of current and upcoming experiments, we chose two representative cases for spectroscopic large scale structure (LSS) surveys, namely the DESI and the \Euclid experiments. We detail their specifications in Table \ref{tab:gal_spec}.

Concerning the CMB, we first considered a \Planck-like experiment, which is currently a state-of-the-art database in terms of multi-frequency, full sky CMB data \citep{Planck2018CosmoParameters}. In order to observe the future sensitivity reachable on the smallest angular scales via ground CMB experiments, we also considered the Simons Observatory \citep{SimonsObs2019} and the CMB Stage 4 \citep{AbazajianAddison2019}. Both cover thousands of square degrees  of the southern sky ($>40\%$ of the sky) with extremely high sensitivity ($\leq 2\,$
~$\mu$K~arcmin) and fine angular resolution (at the arcmin level).

\subsection{The DESI experiment}

In this section, we discuss DESI, which is a ground-based survey that will cover $14,000~\mathrm{deg}^2$ on the sky and will measure the redshift of about 30 million galaxies using optical-fibre spectroscopy \citep{DESI_Collaboration_2016}. It will target four different classes of galaxies. In this work, we computed forecasts for the emission line galaxy (ELG) sample, which is the largest sample of the survey. It ranges from $z=0.6$ up to $z=1.6$.  The expected galaxy distribution $\bar{n}_{\rm g}(z)$ (see Fig.~\ref{fig:nofz_desielg}) and the galaxy bias $b_{\rm}(z)$ are calibrated based on the DEEP2 survey \citep{NewmanCooper2013}. The uneven shape of the redshift distribution of galaxies can be explained by the selection effect of the DESI  survey and by the sample variance in the DEEP2 survey. The (linear) bias of the spatial distribution of this galaxy population with respect to dark matter is a redshift-dependent quantity approximated by
\begin{equation}
    \label{eq:bias_desi}
    b_{\rm g}(z) = 0.84 / D(z) \; ,
\end{equation}
with $D(z)$ denoting the growth factor of linear matter density perturbations.

\subsection{The \Euclid spectroscopic survey}

The \Euclid satellite will observe about $15\,000~\mathrm{deg}^2$ of the extragalactic sky \citep{LaureijsAmiaux2011}. The NISP instruments will provide slitless spectroscopy, allowing for precise redshift determinations for about $1950$~gal~deg$^{-2}$. The spectroscopic survey will target $\rm H_{\alpha}$ emission-line galaxies in the redshift range $0.9<z<1.8$. We assumed model 3 from \cite{PozzettiHirata2016} for the expected number density of galaxies $\bar n_\mathrm{g} (z)$ (see Fig.~\ref{fig:nofz_euclid}).
For the expected galaxy linear bias, we fit a linearly redshift dependent bias on the values of the Table 3 of \cite{EuclidCollaborationBlanchard2019}, yielding
\begin{equation}
    \label{eq:bias_euclid}
    b_\mathrm{g}(z)  = 0.79 + 0.68 \, z \; .
\end{equation}

\subsection{Tomography}

As already mentioned above, our forecasts are based on a tomographic approach where the entire redshift range covered by a galaxy survey is sliced into different redshift bins. Centred at each of these redshift bins, we considered Gaussian redshift shells of a given width $\sigma_{\rm z}$ centred on redshifts $z_i$:
\begin{equation}
    W_i(z) = \exp{\paren{-\frac{(z- z_i)^2}{2\sigma_{\rm z}^2}}} \; .
    \label{eq:GW1}
\end{equation}
Provided that a Gaussian shell dilutes information on radial scales shorter than the Gaussian width, our choice of$\sigma_{\rm z}$ is a compromise between maximising the amount of radial scales under study, and minimising the impact of non-linear, radial scales in the analysis \citep{AsoreyCrocce2012, DiDioMontanari2014}. In HMCMA, we find that, at $z\simeq 1$, down to $\sigma_{\rm z}=0.01,$ the impact of radial non-linearities is either negligible or easily tractable with a Gaussian kernel describing thermal, stochastic, radial motions. We thus adopted $\sigma_{\rm z}=0.01$ for our forecasts.

As shown in \citet{AsoreyCrocce2012}, the angular galaxy clustering analysis can recover the same amount of information as the 3D analysis when the bin size is comparable to the maximum scale probed by the 3D analysis. This gives $\sigmaz \, c / H(z) \simeq 2\pi/k_{\rm max}$, so in our case, for $z = 1$ and $k_{\rm max}=0.2$~$h$~Mpc$^{-1}$ (see Sect.~\ref{sec:observables}), we obtain $\sigmaz \simeq 0.01$, corresponding to our choice of bin size.

For each of the two galaxy surveys under consideration, we took 20 redshift bins, and since the overlap between consecutive bins is not zero, we account for all cross-correlations between shells in the covariance matrix. In this way, redundant information between different shells is fully accounted for. The redshift bins sample the range from $z=0.65$ to $z=1.65$ for DESI, and from $z=0.9$ to $z=1.8$ for \Euclid. These redshift bins are displayed in Figs.~\ref{fig:nofz_desielg} and \ref{fig:nofz_euclid}, together with the expected number density of tracers for each survey.

\subsection{The \Planck experiment}

The \Planck satellite was launched in 2009 and scanned the full sky until 2013 in CMB frequencies.
The satellite hosted two instruments, the HFI operating in six frequency bands between 100 GHz and 857 GHz, and the LFI instrument operating in three bands between 30 GHz and 77 GHz.
The CMB maps were produced by combining these frequencies to remove the contribution from the galaxy and other foreground sources.
The final maps have noise of $27 \, \mu K \,  \rm{arcmin}$,
and an effective beam with a full width at half maximum of 7~arcmin.
The final data release of \Planck was published in \citet{Planck2018CosmoParameters}.

The CMB lensing field has been estimated with a minimum variance quadratic estimator, combining temperature and polarisation data.
It is to date the most precise map of the integral of the density of matter on the full extra galactic sky, covering $\sim70\,\%$ of the sky, which made it possible to obtain an estimate of the lensing-potential power spectrum over lensing multipoles $8\leq L\leq 400$ \citep{PlanckGravLensing2018}.

\subsection{The Simons Observatory}

The Simons Observatory consists of four different telescopes placed in the Atacama desert in Chile, with the goal of providing an exquisite mapping of the CMB intensity and polarisation anisotropies from a few degrees down to arcminute scales. Three of the telescopes have 0.5~m of aperture, and with an angular resolution close to half a degree, map 10~\% of the sky targeting the moderate-to-large angular scales. Their primary goal is to measure large-scale polarisation from the background of primordial gravitational waves.

 Alongside these small telescopes, one 6~m diameter telescope will observe at 27, 39, 93, 145, 225, and 280~GHz, with an angular resolution close to the arcminute, which is necessary to obtain a high-resolution map of the lensing potential of the CMB. It is expected to reach a sensitivity level of $6~\mu$K~arcmin on 40~\% of the sky.

\subsection{CMB Stage 4}

The CMB Stage 4 (CMB-S4) experiment will be the successor of the Simons Observatory and will combine resources with the successor of the South Pole telescope and the BICEP/Keck collaborations.
Its main scope is to measure the imprint of primordial gravitational waves on the CMB polarisation anisotropy, but it will also perform a wide survey with a high resolution that will allow us to probe the secondary anisotropies with unprecedented accuracy.
Its deep and wide survey will cover $\sim 60\%$ of the extragalactic sky and will be conducted over seven years using two 6~m telescopes located in Chile, each equipped with $121,760$ detectors distributed over eight frequency bands from 30 GHz to 270 GHz.
These observations will provide CMB temperature and polarisation maps with a resolution of $\leq 1.5 \, \rm arcmin$ and with a noise level of $1~\mu$K~arcmin. This very high sensitivity at small scales both in temperature and polarisation, on a large fraction of the sky, will ensure an accurate estimation of the CMB lensing potential.


\section{Observables}

\label{sec:observables}

In this paper, we consider three different observables, namely the angular galaxy clustering, the corresponding ARF, and the CMB lensing convergence field. In order to compute the forecasts, we restricted our study to the linear scales, where the cosmological linear theory of perturbations apply.
In practice, we ignored all scales above $k_{\rm max}=0.2$~$h$~Mpc$^{-1}$ at all redshifts. This is a conservative approach, as one could consider a scale cutoff that evolves with redshift as in \citet{DiDioMontanari2014}.
We also assumed that our observables were Gaussian distributed, and that the information content was completely captured by the two-point momenta, and in particular the angular power spectrum, either auto or cross, depending on whether we combined different observables or not.
In what follows, we describe our model of the observables, so that expressions for their angular power spectrum can be derived thereafter.

\subsection{Galaxy angular density fluctuations}
\label{subsec:deltag}

The 3D field of the number density of galaxies is noted as $n_\mathrm{g}(z, \vec{\hat n})$, where $\vec{\hat n}$ denotes a direction on the sky.
The average number density of galaxies at a redshift $z$ is defined by $\bar n_\mathrm{g}(z) = \ave{n_\mathrm{g}(z, \vec{\hat n})}_{\vec{\hat n}}$. The 3D field of galaxy density contrast is then given by
\begin{equation}
    \label{eq:gal_overdens}
    \dg^{\rm 3D}(z, \vec{\hat n}) = \frac{n_\mathrm{g}(z, \vec{\hat n}) - \bar n_\mathrm{g}(z)}{\bar n_\mathrm{g}(z)} \;.
\end{equation}
We assume that the galaxy density contrast traces the dark matter density contrast \dm via a scale-independent bias: $\dg^{\rm 3D}(z, \vec{\hat n}) = b_g(z) \, \delta_m^{\rm 3D}(z, \vec{\hat n})$. This bias depends on the properties of the galaxies used as a tracer for each survey, and they are given in Eqs.~\ref{eq:bias_desi} and \ref{eq:bias_euclid}.

In our analysis, we modelled the observed redshift of galaxies $z_{\rm obs}$ as a 3D field. It is defined as the sum of the redshift induced by the Hubble flow, and the redshift due to the peculiar velocity of galaxies:
\begin{equation}
    z_{\rm obs}(z, \vec{\hat n}) = z + (1+z) \,  \frac{\vec{\mathrm{v}}(z, \vec{\hat n}) \cdot \vec{\hat n}}{c} \; ,
\end{equation}
where $\vec{\mathrm{v}}$ is the peculiar velocity field of galaxies. We neglect other sources of redshift distortions that are significantly smaller than those considered here (HMCMA).

The angular galaxy clustering field is then modelled by an integral along the line of sight in which, at every redshift $z$, only galaxies within the selection function $W(z_{\rm obs}; z_i)$ are included:
\begin{equation}
    \begin{split}
    \dg^i(\vec{\hat n})\ &= \frac{1}{N^i_{\rm g}} \int_{z=0}^{\infty} \;  \diff V_\Omega \; \bar n_\mathrm{g}(z) \, b_\mathrm{g}(z) \, \dm (z, \, \vec{\hat n}) \,
    W_i \brackets{ z_\mathrm{obs}(z, \vec{\hat n}) } ,
    \label{eq:deltag_field}
    \end{split}
\end{equation}

where $N^i_{\rm g} = \int_{z=0}^\infty \diff V_\Omega \; \bar n_\mathrm{g}(z) \,  W_i(z)$ is the average number of galaxies per solid angle, under the $i$-th selection function $W_i$ centred on redshift $z_i$, and in practice can be computed from an angular average over the survey's footprint.

We next expand the selection function, retaining only linear terms in density and velocity fluctuations, finding the following:
\begin{equation}
    \label{eq:deltag_field_dev}
    \begin{split}
        \dg^i(\vec{\hat n})\ &\simeq \frac{1}{N^i_{\rm g}}  \int_{z=0}^\infty \diff V_\Omega\; \bar n_\mathrm{g}(z) \,  W_i(z) \\
         &\times \brackets{b_\mathrm{g}(z) \, \dm (z, \, \vec{\hat n}) + (1+z) \, \frac{\diff \ln W_i}{\diff z} \frac{\vec{\mathrm{v}}(z, \vec{\hat n}) \cdot \vec{\hat n}}{c}}\; ,
    \end{split}
\end{equation}
with the derivative $\diff \ln W_i/ \diff z = - (z - z_i)/\sigmaz^2$.

\subsection{Angular redshift fluctuations}
\label{subsec:deltaz}

The ARF field represents the spatial variations of the average redshift of galaxies on the sky. The average redshift of galaxies is given by
\begin{equation}
    \label{eq:bar_z}
    \begin{split}
    \bar z &= \frac{1}{N^i_{\rm g}} \, \ave{ \int_{z=0}^\infty \diff V_\Omega\; z_{\rm obs}( z, \vec{\hat n})  \,  n_\mathrm{g}(z, \vec{\hat n}) \,W_i \brackets{z_\mathrm{obs}(z, \vec{\hat n})}}_{\vec{\hat n}} \\
    &= \frac{1}{N^i_{\rm g}} \int_{z=0}^\infty \diff V_\Omega \, z\, \bar{n}_\mathrm{g}(z) \, W_i(z)\;.
    \end{split}
\end{equation}

We thus define the ARF field as follows:
\begin{equation}
    \label{eq:dz_field}
    \begin{split}
        \dz^i(\vec{\hat n}) = \frac{1}{N^i_{\rm g}}
         \int_{z=0}^\infty & \diff V_\Omega\; (z_{\rm obs}(z, \vec{\hat n}) - \bar z ) \, \bar{n}_\mathrm{g}(z)\, \\
         &\times   \brackets{1 + b_\mathrm{g}(z) \, \dm (z, \, \vec{\hat n})} \,  W_i \brackets{ z_\mathrm{obs}(z, \vec{\hat n} \, )}
         \; ,
    \end{split}
\end{equation}
where we again refer to a redshift bin centred upon $z_i$.
Expanding the Gaussian selection function at first order and retaining only linear terms in density and velocity, we find
\begin{equation}
    \label{eq:deltaz_model}
    \begin{split}
        \dz^i(\vec{\hat n}) \simeq \frac{1}{N^i_{\rm g}}
        &\int_{z=0}^\infty \diff V_\Omega\; \bar n_\mathrm{g}(z) \, W_i(z) \, \Biggl[ \paren{z- \bar z} \, b_\mathrm{g}(z) \, \dm (z,  \vec{\hat n}) \\
        &\quad + (1+z) \, \frac{\vec{\mathrm{v}}(z, \vec{\hat n}) \cdot \vec{\hat n}}{c} \, \paren{1+ (z-\bar z) \, \frac{\diff \ln W_i}{\diff z}}\Biggr] \; .
    \end{split}
\end{equation}
We note that given the small widths adopted ($\sigma_{\rm z}=0.01$), it is safe to assume that the bias $b(z)$ remains constant within the redshift bin.

\subsection{CMB lensing}
\label{subsec:cmblensing}

The image of the primary CMB, emitted at the moment of recombination at $z\simeq 1100$, is distorted by the gravitational lensing arising as a consequence of the (slightly inhomogeneous) mass distribution between us and the surface of the last scattering. This modifies the initial anisotropy pattern and creates statistical anisotropy \citep[see][for a review]{LewisChallinor2006}. Assuming that the primordial CMB is Gaussian and statistically isotropic, we can reconstruct the lensing potential $\phi$ with the so-called quadratic estimator \citep{HuOkamoto2002,OkamotoHu2003}. The lensing potential is linked to the convergence field \kcmb by
\begin{equation}
    \kcmb = -\frac{1}{2} \Delta \phi \;.
\end{equation}
This convergence field is directly proportional to the surface mass density along the line of sight. The CMB lensing is as such an unbiased estimation of the distribution of mass. However it is an integrated estimation, whereas galaxy surveys can enable tomographic analyses thanks to redshift measurements.

The CMB lensing has been characterised by the \Planck CMB survey \citep{PlanckGravLensing2018}, and by the ACTPol \citep{SherwinvanEngelen2017}, SPT-SZ \citep{OmoriChown2017}, and SPTpol \citep{WuMocanu2019} collaborations.
 Next-generation CMB surveys such as the Simons Observatory or CMB-S4 will increase the signal-to-noise ratios at all $\ell$ by almost one and two orders of magnitude, respectively. These new experiments will  make the CMB lensing a sensitive probe of the dark matter distribution, and via cross-correlation studies it will be crucial to constraining the growth rate of the structure, the neutrino masses, or the level of primordial non-Gaussianities during the inflationary epoch.

\subsection{Angular power spectra}

Our statistical tools to test cosmological models are the angular two-point power spectra $C_\ell$ performed over the three fields defined in Sects. \ref{subsec:deltag}, \ref{subsec:deltaz}, and \ref{subsec:cmblensing}.
Assuming that the galaxy bias and the growth factors are scale independent, one can show that our (cross- and auto-) angular power spectra can be expressed as the convolution of two kernels $\Delta_\ell^\mathrm{A}(k)$ and $\Delta_\ell^\mathrm{B}(k)$, correspondingly for the fields A and B \citep[see, e.g.][]{Huterer2001_powerspec}:
\begin{equation}
    \label{eq:powerspectrum}
    C_\ell^{\mathrm{A}, \mathrm{B}} = \frac{2}{\pi}\int \diff k \; k^2 \, P(k) \, \Delta_\ell^\mathrm{A}(k) \, \Delta_\ell^\mathrm{B}(k),
\end{equation}
where $P(k)$ is the linear 3D matter power spectrum at $z=0$, which is a function of the wave number $k$.

To obtain the theoretical prediction of our angular power spectra, we started from the 2D fields defined in Eqs.~\ref{eq:deltag_field_dev} and \ref{eq:deltaz_model}.
The velocity field is related to the matter density contrast field via the linearised continuity equation $\partial \, \dm /\partial t + \nabla \vec{\mathrm{v}} /a = 0 $, with $a(z)$ being the cosmological scale factor, $a=1/(1+z)$. We introduce the following linear growth rate:
\begin{equation}
   f = \frac{\diff \ln D }{\diff \ln a }= - (1+z) \, \frac{1}{D(z)} \, \frac{\diff D }{ \diff z} \; .
\end{equation}
We assume $f(z) = \Omega_{\rm m}(z)^{\, \gamma}$, with $\gamma=0.55$ \citep{LahavLilje1991, Linder2005}.
The growth factor $D(z)$ is computed by integrating the growth rate $f(z)$.

One can show that the angular galaxy clustering kernel is the sum of two terms, one arising from the density of galaxies and the other from the peculiar line of sight velocities, $\Delta_\ell^{g} = \Delta_\ell^{g}\vert_{\delta} + \Delta_\ell^{g}\vert_{\rm v}$ \citep[see e.g.][]{PadmanabhanSchlegel2007}:
\begin{align}
    \label{eq:deltag_dens_kernel}
    \Delta_\ell^{g,i} \vert_{\delta} (k) &= \frac{1}{N^i_{\rm g}} \int_{z=0}^\infty \diff V_\Omega\; \bar{n}_\mathrm{g}(z) \, W_i(z) \, b_\mathrm{g}(z) \, D(z) \, j_\ell(k\,r(z)) \, ,
    \\
    \label{eq:deltag_vlos_kernels}
    \Delta_\ell^{g,i}\vert_{\rm v}(k) &= \frac{1}{N^i_{\rm g}} \int_{z=0}^\infty \diff V_\Omega\; \bar{n}_\mathrm{g}(z) \, H(z) \, f(z) \, D(z) \, \frac{\diff W_i}{\diff z}\, \frac{j'_\ell(k\,r(z))}{k} \; ,
\end{align}

where $j_\ell (x)$ is the spherical Bessel function of order $\ell,$ and $j'_\ell (x)$ is its derivative $j'_\ell(x)\equiv \diff j_\ell/ \diff x$.

One can thus write the power spectrum as the sum of the contributions from the density and from the velocity kernels $C_\ell = C_\ell^{\, \delta \,, \delta} + 2 \, C_\ell^{\, \delta \,, \rm{v}} + C_\ell^{\, \rm v\, ,v} $.

The ARF kernel can also be separated into two kernels:
\begin{align}
    \label{eq:deltaz_dens_kernel}
    \Delta_\ell^{z,\,i}\vert_{\delta}(k) &= \frac{1}{N^i_{\rm g}} \int_{z=0}^\infty  \diff V_\Omega\; \bar{n}_\mathrm{g}(z) \, W_i(z) \, b_\mathrm{g}(z) \, D(z) \, (z - \bar{z}) \, j_\ell(k\,r(z)) \; ,
    \\
    \label{eq:deltaz_vlos_kernel}
    \begin{split}
        \Delta_\ell^{z,\,i}\vert_{\rm v}(k) &= \frac{1}{N^i_{\rm g}} \int_{z=0}^\infty  \diff V_\Omega\; \bar{n}_\mathrm{g}(z) \,  H(z) \, f(z) \, D(z)  \, W_i(z)\\
        &\qquad \qquad \qquad \times  \left[ 1  + (z - \bar{z}) \, \frac{\diff \ln W_i}{\diff z} \right] \, \frac{j'_\ell(k\,r(z))}{k} \; .
    \end{split}
\end{align}

The kernel function of the CMB lensing convergence field is given by
\begin{equation}
    \label{eq:convergence_kernel}
    \Delta_\ell^\kappa(k) =\frac{3  \Omega_{\rm m,0}}{2} \, \left( \frac{H_0}{c}\right)^2 \int_{r = 0}^{r_*} \diff r \; \frac{r}{a(r)}\, \frac{r_* - r}{r_*} \, D(z(r)) \, j_\ell(k\,r),
\end{equation}
where $r_*$ the comoving distance from the observer to the last scattering surface, and $a$ is the cosmological scale factor.

\begin{figure}
    \centering
    \includegraphics[width=\columnwidth]{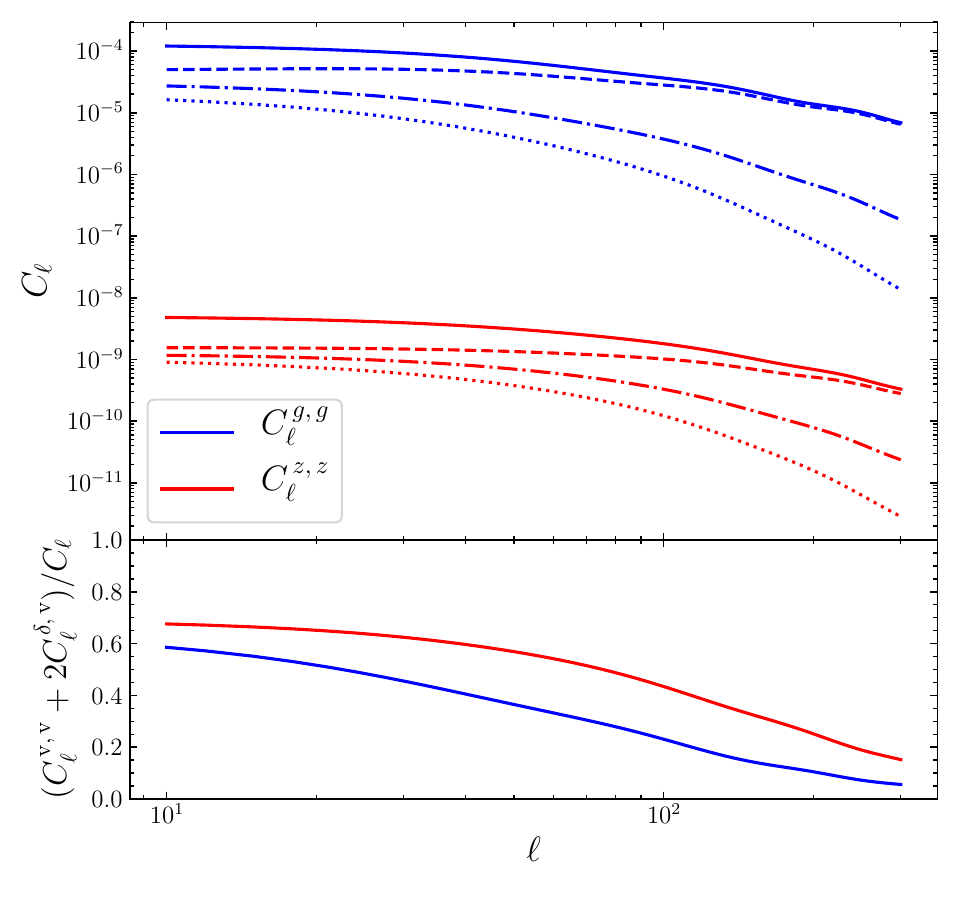}
    \caption{\emph{Top panel}: Power spectra of angular galaxy clustering (\dg, in blue) and ARF (\dz, in red), for a Gaussian redshift bin taken in a DESI-like survey. The bin is centred on $z_i=0.75$ and has a standard deviation of $\sigmaz= 0.01$. The dashed line shows the term coming from the density kernel $ C_\ell^{\, \delta \,, \delta}$ , the dotted line shows the part coming from the velocity kernel $C_\ell^{\, \rm v\, ,v}$ , and the dot-dashed line shows the cross term $C_\ell^{\, \delta \,, \rm{v}}$. The total $C_\ell$ power spectra (plain lines) correspond to the sum $C_\ell = C_\ell^{\, \delta \,, \delta} + 2 \, C_\ell^{\, \delta \,, \rm{v}} + C_\ell^{\, \rm v\, ,v} $. \emph{Bottom panel}: Velocity dependence ratio in the power spectrum ($C_\ell^{\, \rm v\, ,v} + 2 \, C_\ell^{\, \delta \,, \rm{v}}$) over the complete power spectrum, for the angular galaxy clustering (blue line) and for the ARF (red line).  This figure shows that ARF are more sensitive to the peculiar velocity of galaxies than angular galaxy clustering, for the same redshift shell.}
    \label{fig:cl}
\end{figure}

The top panel of Fig.~\ref{fig:cl} shows the angular power spectra of the angular galaxy clustering and ARF for a Gaussian selection function of width $\sigmaz = 0.01$ centred on $z_i=0.75$ in a DESI-like survey. In the same figure, we show the terms arising from the density fluctuation kernel and the peculiar velocity kernel (c.f. Eqs.~\ref{eq:deltag_dens_kernel} to \ref{eq:deltaz_vlos_kernel}).
We can see that the peculiar velocity term is relatively more important (compared to the total power spectrum) in the ARF power spectrum than in the angular galaxy clustering power spectrum.
To better illustrate this fact, in the bottom panel of Fig.~\ref{fig:cl} we show the ratio of the velocity part of the power spectrum (which is the sum $C_\ell^{\, \rm v\, ,v} + 2 \, C_\ell^{\, \delta \,, \rm{v}}$) over the total power spectrum for both angular galaxy clustering and ARF.
For both fields, the peculiar velocity contribution dominates at low $\ell$, while it vanishes to zero  for $\ell>300$.
At $\ell=10$, the velocity-dependent part in the power spectrum represents around $67 \, \%$ of the total contribution for $C_\ell^{\, z, \, z}$, while it represents only $58 \, \%$ of $C_\ell^{\, g, \, g}$.
The difference between the two is even more visible at $\ell=60$, where the velocity contribution represents $55\, \%$ of $C_\ell^{\, z, \, z}$ and only $35\,\%$ of $C_\ell^{\, g, \, g}$.

This difference is caused by the intrinsically different nature of the angular galaxy clustering and ARF transfer functions: angular galaxy clustering is sensitive to the average of density and velocity under the Gaussian shell, whereas ARF is sensitive to radial derivatives of those fields. For narrow shells, this makes both fields practically uncorrelated (HMCMA), and  given the ratio comparison showed in Fig.~\ref{fig:cl}, one would expect ARF to be more sensitive than angular galaxy clustering to cosmological parameters impacting peculiar velocities.


\section{Signal-to-noise forecasts}
\label{sec:ston}

We forecast the expected signal-to-noise ratio for different combinations of observables.
Our data vector $\vec{D}(\ell)$ contains the auto- and cross-power spectra between the different observables and between the redshift bins. In order to compare several combinations of probes, we define the following data vectors:
\begin{align}
    &\Dg(\ell) = \paren{C^{\,g_i, \, g_j}_l} \; ,  \label{eq:datavec0} \\
    &\Dz(\ell) = \paren{C^{\,z_i, \, z_j}_l} \; \nonumber,\\
    &\Dgz(\ell) = \paren{C^{\,g_i, \, g_j}_l, C^{\,g_i, \, z_j}_l, C^{\,z_i, \, z_j}_l} \; ,\nonumber \\
    &\Dgk(\ell) = \paren{C^{\,g_i, \, g_j}_l, C^{\,g_i, \, \kcmb }_l, C^{\,\kcmb, \, \kcmb }_l}\; , \nonumber \\
    &\Dgzk(\ell) = \biggl( C^{\,g_i,\,  g_j}_l, C^{\,g_i,\,  z_j}_l,  C^{\,g_i,\,  \kcmb }_l, C^{\,z_i,\,  z_j}_l,  \nonumber \\
    & \phantom{xxxxxxxxxxxxxxxxxxxxxx} C^{\,z_i, \, \kcmb }_l, C^{\,\kcmb ,\,  \kcmb }_l \biggr) \, , \label{eq:datavec1}
\end{align}
where $i$ and $j$ are indexes running over the redshift bins.
We performed a tomographic analysis with 20 redshift bins, thus the data vectors containing only the auto-spectra of angular galaxy clustering and ARF (\Dg and \Dz) contain 210 $C_\ell$ each. The data vector containing the cross-correlation \Dgz has 820 $C_\ell$ and the longest data vector \Dgzk contains 861 $C_\ell$.

\begin{figure}
    \centering
    \includegraphics[width=\columnwidth]{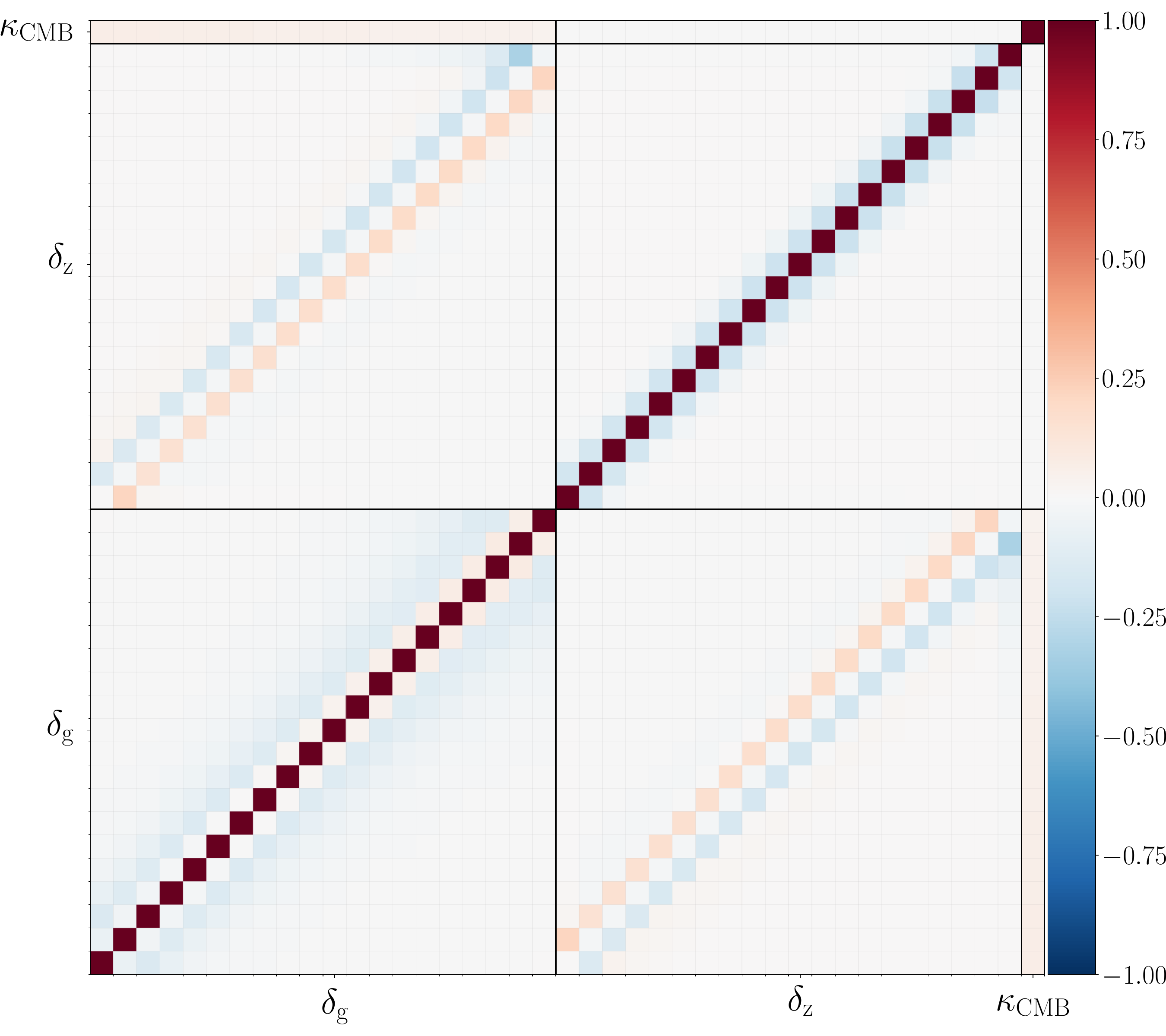}
     \caption{Correlation matrix between our observables for the 20 redshift bins in an \Euclid-like survey, at $\ell=10$. This matrix corresponds to the $\Dgzk(\ell=10)$ data vector. The value in each pixel corresponds to $C_\ell^{\rm A, B} / \sqrt{C_\ell^{\rm A, A} C_\ell^{\rm B, B}}$.
     All the data vectors considered in Eqs.~\ref{eq:datavec0} to \ref{eq:datavec1} are a subset of this matrix.
    We see that there is no correlation between \dg and \dz inside the same redshift bin (diagonals of the upper-left and lower-right blocks), and that there are opposite and positive correlations for neighbouring bins.}
    \label{fig:correlation_cl_euclid}
\end{figure}

In Fig.~\ref{fig:correlation_cl_euclid}, we display the correlation matrix for the $\Dgzk(\ell=10)$ data vector. We clearly see that, in the same redshift bin, angular galaxy clustering and ARF are practically un-correlated (diagonal terms of the top-left and lower-right blocks close to zero), but that there is some degree of anti-correlation in neighbouring redshift bins.
We can also observe that the CMB lensing field is almost uncorrelated with the ARF.

We assume that there is no correlation between different multipoles and that the covariance between the probes is totally captured by a Gaussian covariance. This assumption is exact on large (linear) scales and if the survey covers the full sky. With regard to real data, the footprint of the survey and the presence of masked area will create correlations between multipoles.
We invite the reader to consult, for example, \citet{KrauseEifler2017} or \citet{Lacasa2018} for the inclusion of the higher order (non-Gaussian) terms in the covariance matrix.
In this work, we neglect these effects.

\begin{figure}
    \centering
    \includegraphics[width=\columnwidth]{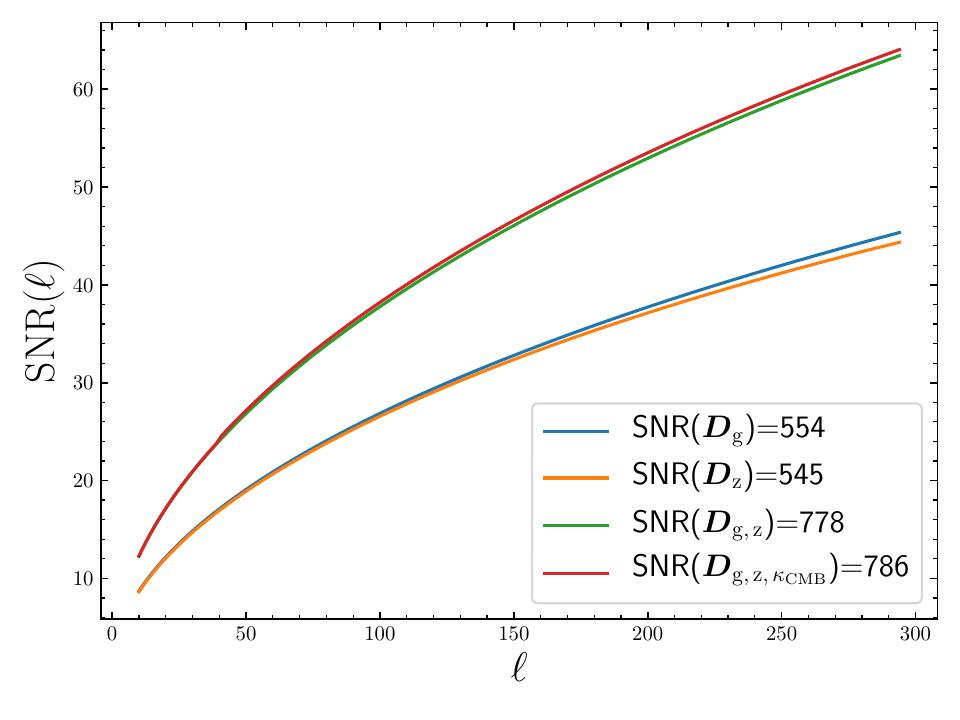}
     \caption{Signal-to-noise ratios of angular galaxy clustering (\Dg) in blue, ARF (\Dz) in orange and the combinations \Dgz in green and \Dgzk in red. We used 20 tomographic Gaussian bins of $\sigma_\mathrm{z}=0.01$ in width in an \Euclid-like survey, in combination with a CMB-S4 survey. The total signal-to-noise ratios for the range of multipoles $\ell=10$ to $\ell=300$ are shown in the text box on the bottom right.}
    \label{fig:ston}
\end{figure}

The signal-to-noise ratios of our data vectors as a function of $\ell$, taking into account all redshift bins and the correlations between them, are given by
\begin{equation}
    \mathrm{S/N}\paren{\vec{D}(\ell)} = \sqrt{\vec{D}(\ell)^{\, t} \, \tens{Cov}_\ell^{-1} \, \vec{D}(\ell) } \; ,
\end{equation}
and the total signal-to-noise ratios are
\begin{equation}
    \mathrm{S/N}\paren{\vec{D}} = \sqrt{\displaystyle \sum_{\ell =\ell_{\rm min}}^{\ell_{\rm max}} \brackets{ \mathrm{SNR}\paren{\vec{D}(\ell)}}^2 } \; .
\end{equation}

Assuming that there is no correlation between different multipoles, we defined our Gaussian covariance matrix between our data vectors as in \citet{HuJain2004}:
\begin{equation}
    \label{eq:covariance}
    \begin{split}
    \tens{Cov}_\ell\left(C_{\ell}^{\,\rm{A}, \rm{B}}, C_{\ell}^{\,\rm{C}, \rm{D}} \right) &=
        \frac{1}{(2\ell + 1) \, \Delta\ell \, f_{\rm sky}} \\
        & \times \left[ \left(C_{\ell}^{\, \rm{A}, \rm{C}} + \delta^{\rm K}_{\rm A, C} \, N^{\,\rm{A}}_{\ell}\right)
        \left(C_{\ell}^{\, \rm{B}, \rm{D}} + \delta^{\rm K}_{\rm B, D} \, N^{\,\rm{B}}_{\ell}\right) \right. \\
        &+ \left.
        \left(C_{\ell}^{\,\rm{A}, \rm{D}} + \delta^{\rm K}_{\rm A, D} \,N^{\,\rm{A}}_{\ell}\right)
        \left(C_{\ell}^{\,\rm{B}, \rm{C}} + \delta^{\rm K}_{\rm B, C} \, N^{\,\rm{B}}_{\ell}\right) \right] \; ,
        \end{split}
\end{equation}
with A, B, C, D being the observables $\curly{g_i, \, z_j, \, \kcmb}$, $\Delta\ell$ the width of the multipole bin, $\delta^K_{x,y}$ the Kronecker delta, $N_\ell$ the probe-specific noise power spectra, and $f_{\rm sky}$ the sky fraction of the survey considered.

For the sake of simplicity, when combining galaxy surveys with CMB lensing, we always assume a full overlap of the two. As such, the sky fraction $f_{\rm sky}$ is always taken to be the one of either DESI or \Euclid. Even if not accurate, this provides a rough estimate of the available constraining power that the combination of galaxy surveys with CMB lensing will be able to achieve.

We assumed that the noise of the angular galaxy clustering and that of the ARF were the shot noises arising from the discrete nature of galaxy surveys. We modelled it by replacing the power spectrum of dark matter by a Poissonian term, $P_{\mathrm{shot}}(k,z) = 1 / \bar{n}_{\mathrm{g}}(z)$, in Eq.~\ref{eq:powerspectrum}. From this, we can derive the following expressions for the shot noise:
\begin{align}
    N_\ell^{\, g_i, \, g_j} &= \frac{\delta^{\rm K}_{i,\,j}}{N_\mathrm{gal}^i} \; , \\
    N_\ell^{\, z_i, \, z_j} &= \frac{\delta^{\rm K}_{i,\, j}}{\paren{N_\mathrm{gal}^i}^2} \int \diff V_\Omega\; \bar{n}_\mathrm{g}(z) \, W(z_i,z) \, \left( z - \bar{z}_i \right)^2 \; , \\
    N_\ell^{\, g_i, \, z_j} &= \frac{\delta^{\, i}_{\, j}}{\paren{N_\mathrm{gal}^i}^2} \int \diff V_\Omega\; \bar{n}_\mathrm{g}(z) \, W(z_i,z) \, \left( z - \bar{z}_i \right)  = 0 \; .
\end{align}
We can see here that the shot noise cancels out when computing the cross-correlation between the angular galaxy density and the ARF fields.

The noise of the CMB lensing field reconstructed from \Planck is taken from \cite{PlanckGravLensing2018}. For the forecasted Simons Observatory CMB lensing noise, we took the publicly available noise curves provided by \citet{SimonsObs2019}.\footnote{We used version 3.1.0 of the noise curves available at \url{https://github.com/simonsobs/so_noise_models}.} In practice, we used the noise curves obtained with the internal linear combination (ILC) component separation method, assuming the baseline analysis for a sky fraction of $f_\mathrm{sky} =0.4$. For CMB-S4, the lensing noise curve is taken as the minimum variance N0 bias, which is computed using the code \texttt{quicklens.}\footnote{\url{https://github.com/dhanson/quicklens}} We assume that CMB-S4 will have a beam size (full width at half maximum) of $1 \, \rm arcmin$, a temperature noise of $\Delta T =  1 \, \mu \rm  K \,  arcmin, $ and a polarisation noise of $\Delta P =  \sqrt{2} \, \mu \rm  K \,  arcmin $ \citep{AbazajianAddison2019}.

For both the Simons Observatory and CMB-S4, $\ell=40$ is the minimum multipole that will be accessible. We assume that these measurements will be combined with the \Planck lensing signal for lower multipoles.
As a result, we used the lensing noise of \Planck for mutipoles below $\ell=40$ when forecasting constraints with the Simons Observatory and CMB-S4.

We used the linear matter power spectrum $P(k)$ computed with the \texttt{CLASS} software \citep{Class}. In order to focus on the linear regime we restrict our analysis to a maximum multipole of $\ell_{\rm max}=300$. Assuming the Limber approximation $k=(\ell+1/2)/\chi(z)$, this $\ell_{\rm max}$ corresponds to $k=0.18~h~\rm{Mpc}^{-1}$ at a redshift of $z=0.65$. Given that we sample higher redshifts, we probe larger scales ($k$ lower than $0.18~h~\rm{Mpc}^{-1}$). We therefore expect little impact from non-linear physics in our observables (these are expected to become relevant on $k < 0.2~h~\rm{Mpc}^{-1}$ at $z=0$, and yet shorter at higher redshifts).
Again, this is a conservative approach as one could consider a multipole cuttoff evolving with redshift as in \citet{DiDioMontanari2014}.
We stress that, in our computations, we did not use the Limber approximation but the full computation of spherical Bessel functions.

Our minimum multipole was chosen to be $\ell_{\rm min} = 10$.
To reduce numerical noise and to speed up Fisher matrix computations, we performed a linear binning of the multipoles.
In each multipole bin $[\ell_i, \ell_{i+1}[$, the binned $C_\ell$ is the average of the $C_\ell$'s that fall in the bin, and the binned multipole was taken as $\ell = (\ell_i + \ell_{i+1})/2$.
We chose a bin size of $\Delta_\ell=3$, which was applied to the full $\ell$ range. We checked that this binning did not impact the constraints from the Fisher matrix by comparing it with the case where we did not perform any  multipole binning.

In Fig.~\ref{fig:ston}, we show the signal-to-noise ratio for an \Euclid-like survey combined with a CMB-S4 survey, for four probe combinations of: \Dg, \Dz, \Dgz, and \Dgzk following the redshift binning shown in Fig.~\ref{fig:nofz_euclid}.
The total signal-to-noise for these four data vectors is, respectively, 544, 545, 778, and 786. This shows that the tomographic analysis of angular galaxy clustering and ARF have a similarly high signal-to-noise ratio. Moreover, the combined analysis \Dgz brings more information than measuring the angular galaxy clustering alone \Dg, as the signal-to-noise ratio is increased by $40\%$.

\section{Fisher forecasts}
\label{sec:fisher}

We used the Fisher formalism to compute, \emph{\emph{a priori,}} how well our data vectors defined in Sect.~\ref{sec:ston} will constrain cosmological parameters in the context of future surveys.
As we assumed that there is no correlation between different multipoles, the Fisher matrix can be summed over the multipoles and is given by
\begin{equation}
    \label{eq:fisher}
    F_{\,i,\,j} = \sum_{\ell_{\rm min}}^{\ell_{\rm max}} \; \frac{\partial \vec{D}(\ell)}{\partial \lambda_i} \, \tens{Cov}_\ell^{-1} \, \frac{\partial \vec{D}(\ell)}{\partial \lambda_j},
\end{equation}
with $\vec{D}$ being one of the data vectors defined in Eqs.~\ref{eq:datavec0} to \ref{eq:datavec1}, $\left\{\lambda_i\right\}_i$ the set of free parameters of our model, and $\tens{Cov}_\ell$ the covariance matrix given in Eq.~\ref{eq:covariance}.

The derivatives  $\partial \vec{D}(\ell) / \partial \lambda_i$ are computed as the two-point variation with a 1\% step around the fiducial value. We checked that our derivatives were numerically stable when changing the step size.

We computed forecasts for two cosmological models. The first one assumes the standard \lcdm model, and the parameters we vary are $\left\{ \Omega_{\rm m}, \Omega_{\rm baryon}, \sigma_8, n_\mathrm{s}, h,\right\}$. The fiducial values of these parameters are given by \cite{Planck2018CosmoParameters}.
The second model assumes an evolving dark energy equation of state, with the so-called CPL parametrisation \citep{Chevallier2001,Linder2003}: $w(z) = w_0 + w_\mathrm{a} \, z/(1+z)$.
Our second set of free parameters is then $\left\{ \Omega_{\rm m}, \Omega_{\rm baryon}, \sigma_8, n_\mathrm{s}, h, w_0, w_\mathrm{a} \right\}$. In both cases, we assumed a flat universe ($\Omega_k = 0$) with massless neutrinos ($\sum m_\nu = 0$). In Tab.~\ref{tab:fiducial}, we show the fiducial values of the free parameters.

We also considered a bias parameter assumed constant within each redshift bin, thus adding one free parameter for each redshift shell, over which we marginalised the Fisher analysis. The fiducial values of the galaxy bias depend on the survey considered and are given in Eqs.~\ref{eq:bias_desi} and \ref{eq:bias_euclid}. We took the value at $z_i$, which is the centre of the Gaussian shell for each bin.

\begin{table}
    \centering
    \begin{tabular}{c c c c c c c}
        \hline
        \hline
        $\Omega_{\rm b}$ & $\Omega_{\rm m}$ & $n_{\rm s}$ & $h$ & $\sigma_8$ & $w_0$ & $w_{\rm a}$ \\
        \hline
        0.04897 & 0.3111 & 0.9665 & 0.6766 & 0.8102 & -1 & 0 \\
        \hline
    \end{tabular}
    \caption{Fiducial values of the free parameter of our fiducial cosmological model. We first consider only parameters in the standard \lcdm model, and later we include the $w_0$, $w_\mathrm{a}$ parameters from the CPL parametrisation of dark energy.}
    \label{tab:fiducial}
\end{table}

\subsection{Results for the \lcdm model}

\begin{figure}
    \centering
    \includegraphics[width=0.5\textwidth]{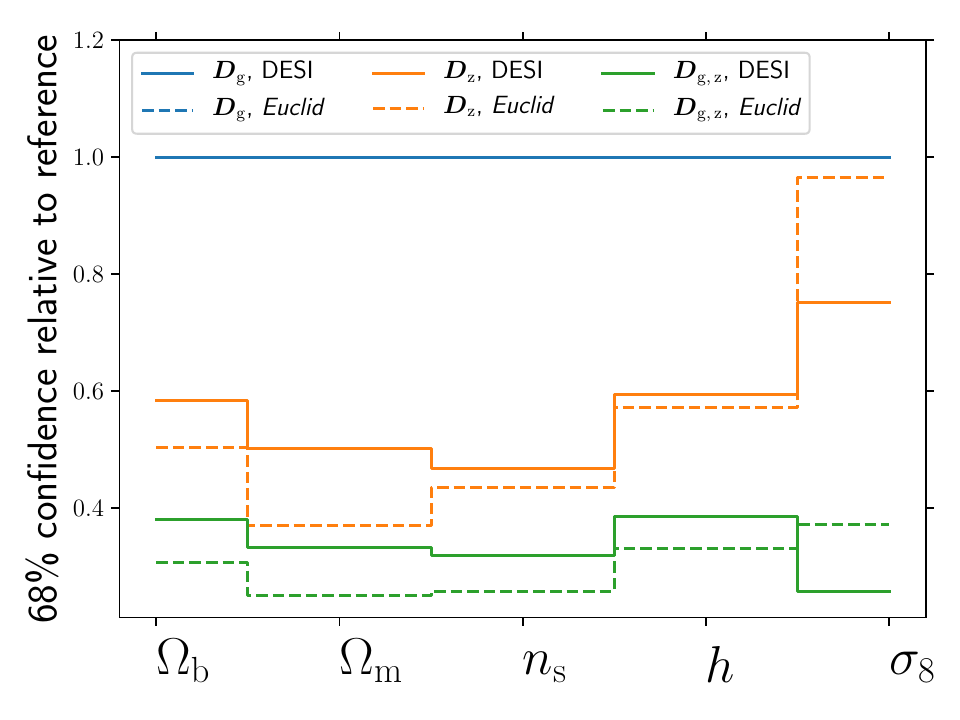}
    \caption{Ratio of  $1\sigma$ confidence interval relative to the $1\sigma$ value from angular galaxy clustering (\Dg) for \lcdm parameters. Constraints are marginalised over the 20 galaxy bias parameters. Plain lines are for a DESI-like survey, while dashed lines are for an \Euclid-like survey. Blue line shows \Dg  (our reference here), orange lines show \Dz, and green lines show \Dgz. We see that for most parameters (except $\sigma_8$) confidence intervals shrink by $\sim50\%$ when using \Dz instead of \Dg. When using the combination \Dgz, $1\sigma$ intervals are shrunk by at least $60\%$ for all parameters.}
    \label{fig:margerr_desi_euclid_lcdm}
\end{figure}

\begin{figure}
    \centering
   \includegraphics[width=0.5\textwidth]{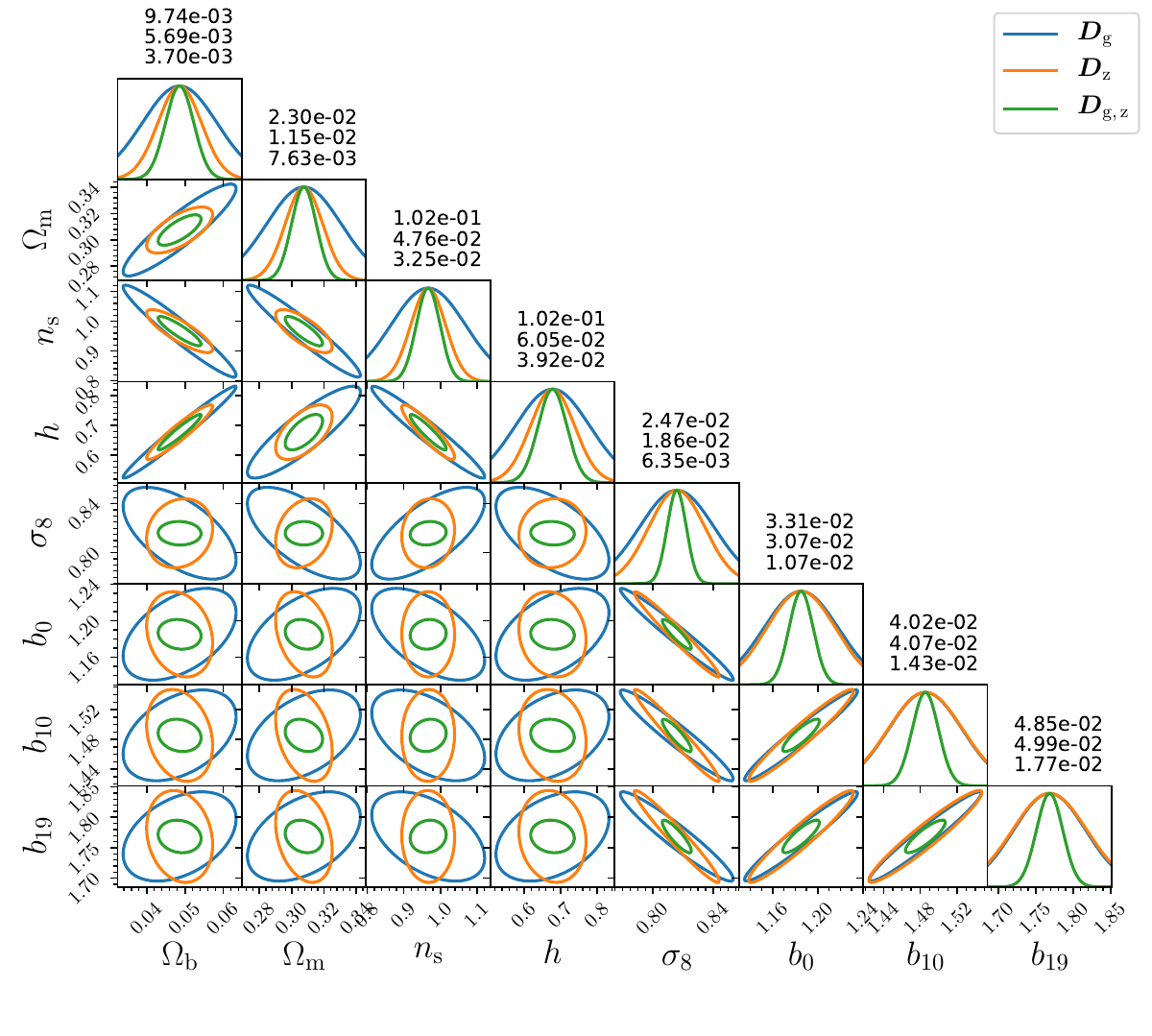}
    \caption{Foreseen constraints ($1\sigma$ contours) for a set of five \lcdm parameters, plus three galaxy bias parameters (out of a total of 20) for a DESI-like survey. We assume 20 tomographic Gaussian bins of size $\sigmaz=0.01$. The blue lines are the constraints for angular galaxy clustering alone \Dg, the orange lines are for the ARF alone \Dz, and the green line is a joint analysis of both fields \Dgz. The figures above the 1-D PDFs give the marginalised $1\sigma$ uncertainty of the parameter for each data vector. We show here only three galaxy bias parameters, even if we marginalised upon the 20 bias parameters.}
    \label{fig:ellipses_LCDM_DESI}
\end{figure}

\begin{figure}
    \centering
    \includegraphics[width=0.5\textwidth] {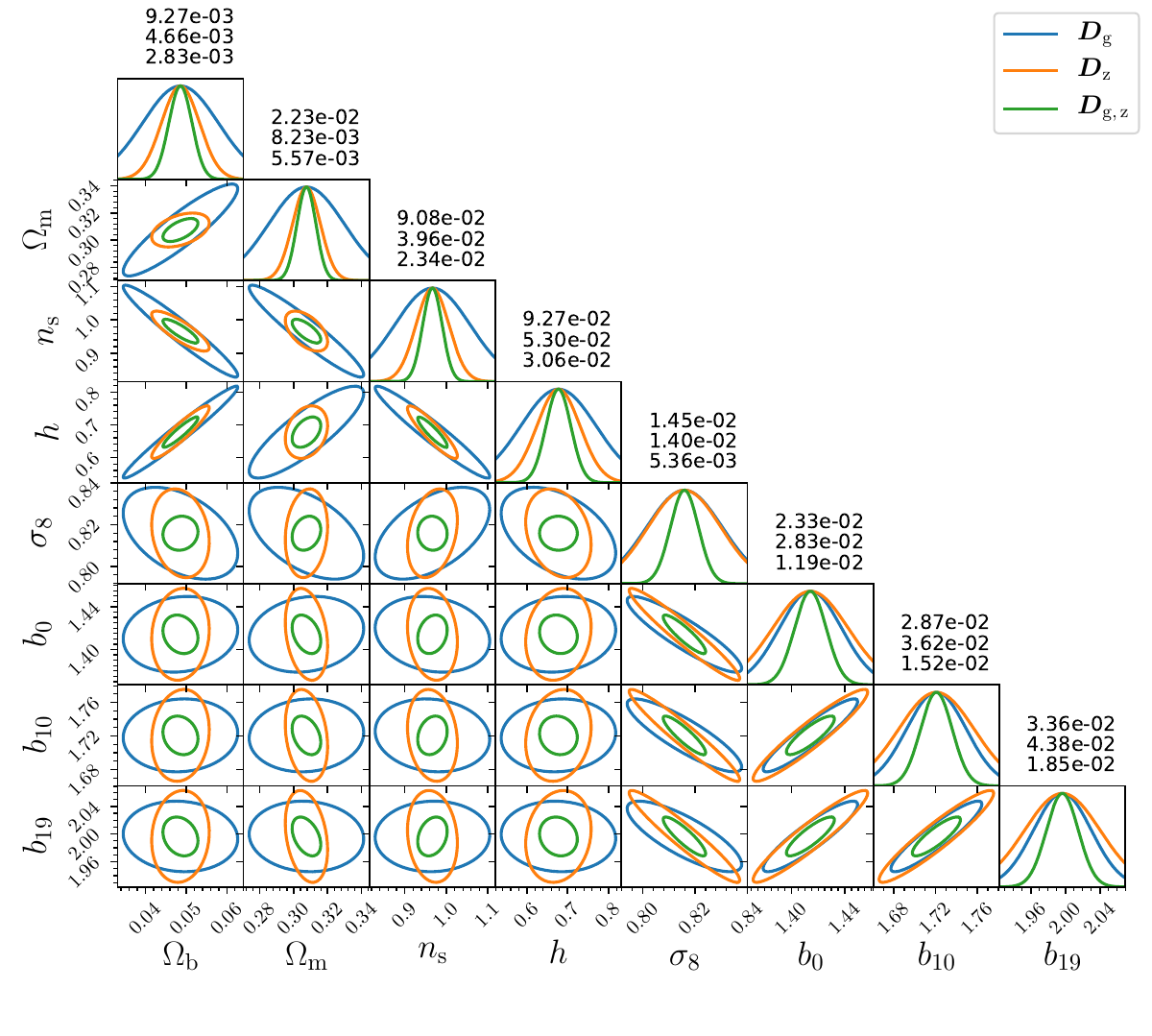}
    \caption{Same as Fig.~\ref{fig:ellipses_LCDM_DESI}, but for an \Euclid-like survey.}
    \label{fig:ellipses_LCDM_Euclid}
\end{figure}

The results for the \lcdm model are summarised in Fig.~\ref{fig:margerr_desi_euclid_lcdm}, where we show the ratio of the $1\sigma$ marginalised uncertainties when including ARF compared to using only angular galaxy clustering, for a DESI-like and a \Euclid-like surveys.
Figures 6 and~\ref{fig:ellipses_LCDM_Euclid} show the $1\sigma$ uncertainty ellipses for the \lcdm parameters and three out of the 20 galaxy bias parameters for a DESI-like survey and an \Euclid-like survey, respectively.
Error ellipses for \Dg, \Dz, and \Dgz are given by blue, orange, and green curves, respectively, while marginalised 1$\sigma$ uncertainties for each parameter are quoted, for these three sets of observables, above the panels containing the 1D probability density distributions (PDFs).

For both types of LSS surveys, we can see in Fig.~\ref{fig:margerr_desi_euclid_lcdm} that ARF (\Dz) are significantly more sensitive than angular galaxy clustering (\Dg), reducing the marginalised uncertainties of all cosmological parameters by a factor of two, except for $\sigma_8$, to which both observables are similarly sensitive.
For the combined analysis \Dgz, marginalised uncertainties are reduced by more than $60 \,\%$ for all parameters (including $\sigma_8$), compared to the angular galaxy clustering probe alone \Dg.
We find that using ARF in combination with angular galaxy clustering provides almost the same improvement on the constraints on cosmological parameters for both surveys, although the improvement is on average slightly better for our \Euclid-like survey.

We see in Figs.~\ref{fig:ellipses_LCDM_DESI} and \ref{fig:ellipses_LCDM_Euclid} that while the degeneracy direction between different cosmological parameter pairs seems very similar for both angular galaxy clustering and ARF, this is again different for $\sigma_8$. For \Dz, this parameter seems rather independent of other cosmological parameters, while its degeneracy with bias parameters is slightly tilted with respect to that of \Dg. As a consequence, the joint \Dgz ellipses show little degeneracy with other parameters, including bias.
We also find that the marginalised constraints from both experiments are very close, although the \Euclid-like experiment provides slightly more sensitive forecasts.

Figure \ref{fig:param_corr_lcdm_desi_full} in the appendix shows the correlation matrix between our free parameters (including galaxy bias parameters) and illustrates the opposite degeneracies that both $\sigma8$ and bias parameters have with the other parameters when comparing ARF and angular galaxy clustering.

Even for those parameters for which both angular galaxy clustering and ARF show a similar direction of degeneracy, the combination of the two observables yields significantly reduced error ellipses. This is mostly due to the lack of correlation between the ARF and angular galaxy clustering for narrow widths used in this work ($\sigmaz \leq 0.01$), as noted in HMCMA and shown here in Fig.~\ref{fig:correlation_cl_euclid}.

\subsection{Extension to CPL dark energy parametrisation}

\begin{figure}[h]
    \centering
    \includegraphics[width=0.5\textwidth]{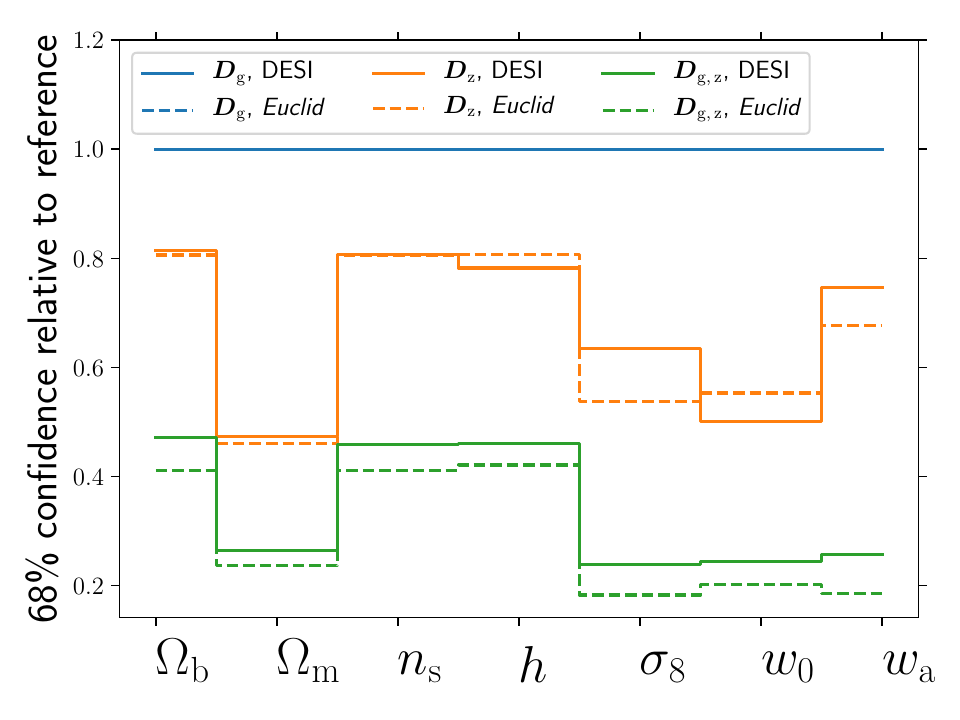}
    \caption{Ratios of $1\sigma$ marginalised uncertainties relative to 1$\sigma$ marginalised uncertainty for \Dg. We assume a \wcdm model and marginalise on 20 galaxy bias parameters (one for each redshift bin). Orange lines show the ratio for \Dz and green lines show the ratio for \Dgz. Solid lines are for a DESI-like survey, while dashed lines are for an \Euclid-like survey. We see that \Dz improves constraints by up to $50\,\%$ compared to \Dg, and the combined analysis \Dgz improves constraints by up to $80\,\%$.}
    \label{fig:margerr_desi_euclid_wcdm}
\end{figure}

We repeat the analysis detailed above including two new parameters describing the equation of state of dark energy following the CPL parametrisation: $w_0$ and $w_{\rm a}$.
In Fig.~\ref{fig:margerr_desi_euclid_wcdm}, we show the improvement on the marginalised uncertainties of the ARF with respect to angular galaxy clustering alone.
We see that \Dz improves the constraints by $20\,\%$ to $50\,\%$ on this set of free \wcdm parameters, for both surveys. The combined analysis \Dgz reduces the uncertainties by at least $50\,\%$ and up to $80\,\%$ for $\Omega_{\rm m}$, $\sigma_8$, $w_0$ and $w_{\rm a}$.
The error ellipses are given in the appendix in Fig.~\ref{fig:ellipses_w0wa} for the DESI-like and \Euclid-like experiments, displaying a pattern similar to what was found for \lcdm, together with the correlation matrices (Fig.~\ref{fig:param_corr_w0wa_desi}).

\begin{figure}[h]
    \centering
    \includegraphics[width=0.5\textwidth]{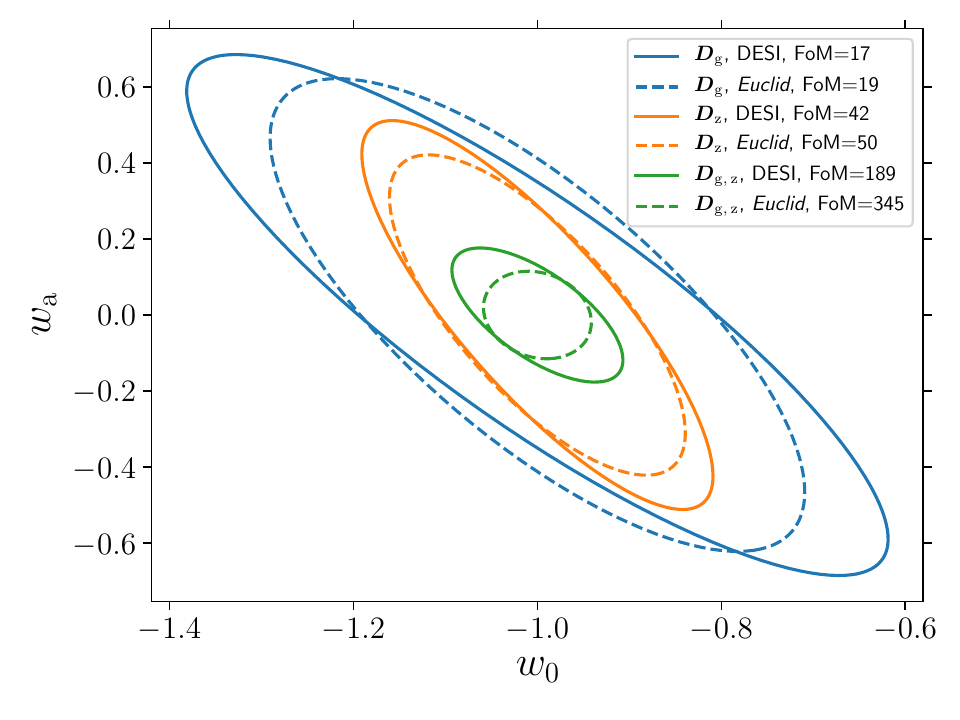}
    \caption{Marginalised constraints ($1\sigma$ contours) on the dark energy equation of state parameters for the DESI-like (solid lines) and for the \Euclid-like (dashed lines) surveys, assuming 20 tomographic Gaussian bins of $\sigmaz=0.01 $. The blue lines are the constraints for angular galaxy clustering alone, the orange lines are for ARF alone, and the green lines are a joint analysis of both fields, \Dgz. These contours are marginalised over the set of cosmological parameters as before, and over the galaxy bias in the 20 redshift bins. We display the figure of merit (FoM) of this pair of parameters in the upper right box for each combination of observables and for each survey.}
    \label{fig:ellipse_w0wa}
\end{figure}

In our idealised case, the combination of ARF with angular galaxy clustering greatly improves the sensitivity of these surveys to dark energy.
As shown in Fig.~\ref{fig:ellipse_w0wa}, the figure of merit of $w_0$-$w_{rm a}$ increases by more than a factor of ten when ARF are combined with angular galaxy clustering.
It increases from 17 to 189 for our DESI-like survey and from 19 to 345 for our \Euclid-like survey.

\subsection{Combining ARF and galaxy clustering with CMB lensing}

\begin{figure}
    \centering
    \includegraphics[width=0.45\textwidth]{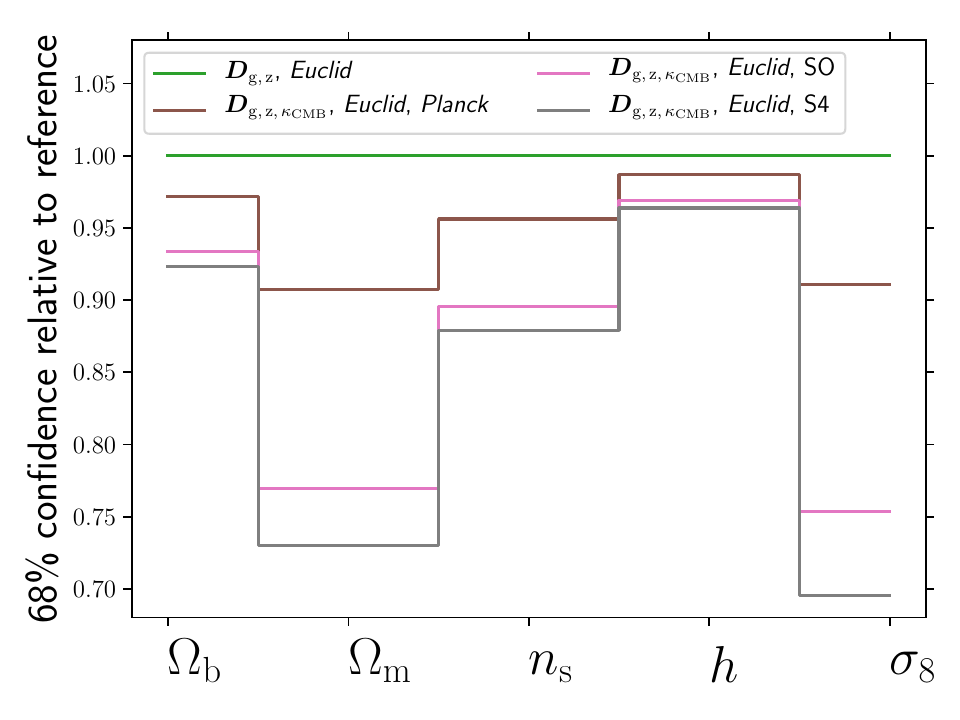}

    \caption{Ratio of $1\sigma$ constraints for \lcdm parameters  from \Dgzk over the $1\sigma$ constraints from \Dgz for the \Euclid-like spectroscopic survey. We show combinations with CMB lensing from \Planck (brown), the Simons Observatory (pink), and CMB-S4 (grey). Constraints are marginalised over the 20 galaxy bias parameters.}
    \label{fig:marg_err_euclid_cmblensing_lcdm}
\end{figure}

\begin{figure}
    \centering

    \includegraphics[width=0.45\textwidth]{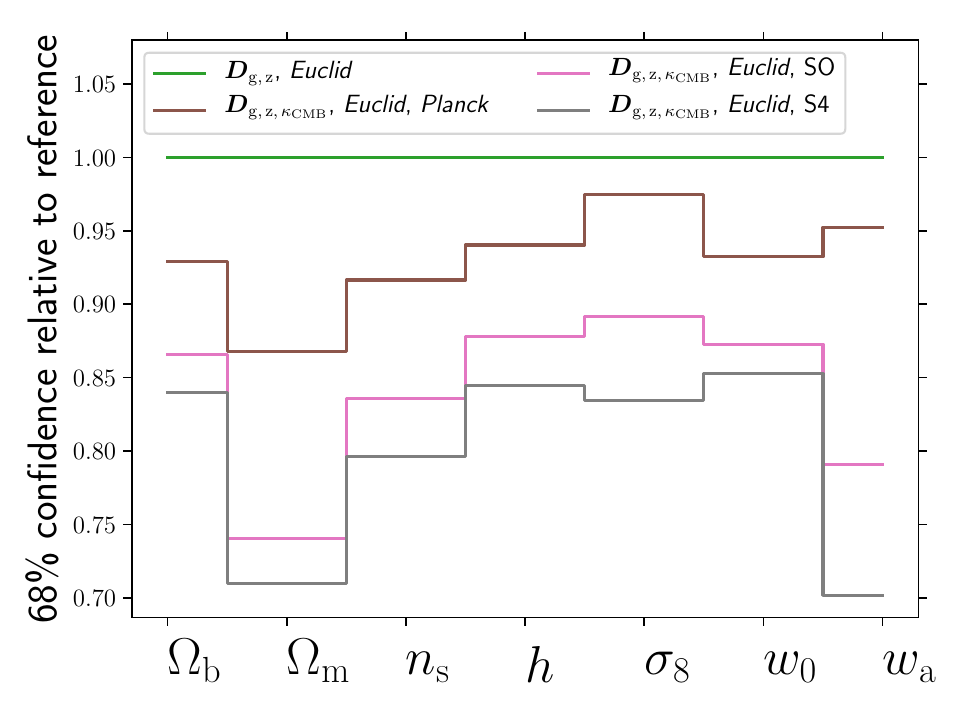}
    \caption{Same as Fig.~\ref{fig:marg_err_euclid_cmblensing_lcdm} for \wcdm parameters.}
    \label{fig:marg_err_euclid_cmblensing_wcdm}
\end{figure}

In Figs 10 and~\ref{fig:marg_err_euclid_cmblensing_wcdm}, we show the improvements on the constraints of the \lcdm and \wcdm  parameters for an \Euclid-like survey, when combined with CMB lensing from \Planck, the Simons Observatory, and CMB-S4, marginalised over the galaxy bias parameters.
We see that including CMB lensing from \Planck improves the constraints by maximum of $10\, \%$ in both cosmologies.
The improvement is more significant when combining ARF and galaxy clustering with the Simons Observatory or CMB-S4. For the Simons Observatory and CMB-S4, in the \lcdm model, marginalised uncertainties on $\Omega_{\rm m}$  and $\sigma_8$ are decreased by up to $30 \, \%$. Other parameters are improved by $5\,\%$ to $10\,\%.$
For the \wcdm model, the improvement is of $\sim 15 \, \%$ for most parameters, with the most significant for $\Omega_{\rm m}$ and $w_{\rm a}$, with uncertainties decreased by up to $30 \, \%$.
We see that the combination with CMB lensing helps to decrease uncertainties on the \wcdm cosmology.

\begin{figure}
    \centering
    \includegraphics[width=0.45\textwidth]{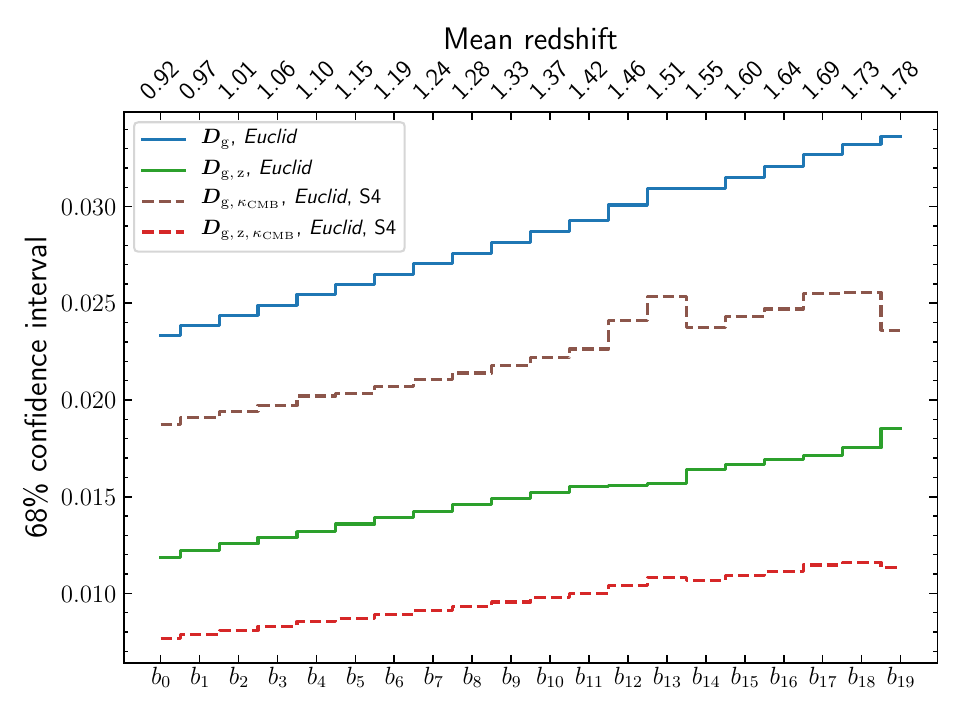}
    \caption{Marginalised 1$\sigma$ confidence values for galaxy bias parameters, with an \Euclid-like survey alone (plain lines) or in combination with the CMB-S4 lensing survey (dashed lines). We marginalised the five free parameters of the \lcdm model. We show constraints with angular galaxy clustering (blue and brown lines) and in combination with ARF (green and red lines).
    The mean redshift of each shell is shown at the top.
    We see that the ARF combined with angular galaxy clustering \Dgz provides better constraints on galaxy bias than the combination of angular galaxy clustering with CMB lensing \Dgk.}
    \label{fig:margerr_bias_euclid_cmblensing}
\end{figure}

Since the CMB lensing is an unbiased probe of the distribution of matter, one of the main interests of combining it with galaxy surveys is to produce tight constraints on the galaxy bias parameter.
In Fig.~\ref{fig:margerr_bias_euclid_cmblensing}, we show the $1\sigma$ marginalised uncertainties on the galaxy bias parameters for each of the 20 redshift bins in an \Euclid-like survey combined with CMB-S4 lensing, for the \lcdm model.
We compared the constraints obtained for angular galaxy clustering alone (\Dg), with the ones obtained when combined with CMB lensing (\Dgk), with ARF (\Dgz), and then the full combination (\Dgzk).

We see that the combination of angular galaxy clustering with ARF provides better constraints on the galaxy bias than the combination with CMB lensing.
For instance, at a redshift of 1.06, the marginalised uncertainties for the galaxy bias parameter $b_3$ is of 0.025 for the angular galaxy clustering, it decreases to 0.020 when combined with CMB lensing, and down to 0.013 when combined with ARF. The combination of the three results in marginalised uncertainties of 0.08.
We can see that the CMB lensing improves constraints by $\sim20\,\%$ only, while ARF improves constraints by $\sim50\,\%$ (a factor of two improvement).
We argue that this is due to the importance of the velocity term in the ARF kernel (see Fig.~\ref{fig:cl}), which does not depend on galaxy bias as it is sensitive to the full matter distribution.

\section{Discussion}
\label{sec:discussion}

One could argue several reasons why angular observables might be preferred over standard 3D ones.
Probably the main one is the lack of assumption of any fiducial cosmological model to analyse the data.
This means that angular observables may be {\em \emph{directly}} compared with theoretical predictions without any intermediate data manipulations that hinge on an assumption of which the implications in the analysis may not always be clear.
Moreover, this type of angular analysis is conducted tomographically in moderately narrow redshift shells, thus avoiding the assumption that the universe remains effectively frozen in relatively long time spans, as it may occur in 3D clustering analysis where an {\em \emph{effective}} redshift must be defined for the entire volume under analysis \citep[see, e.g.][]{cuesta_BAO_dr12}.
\citet{AsoreyCrocce2012} and \citet{DiDioMontanari2014} have shown that when using a large number of narrow redshift slices, a 2D clustering analysis can produce the same constraints on cosmological parameters as a 3D clustering analysis, provided that the width of the redshift slices is comparable to the minimum scale probed in the 3D analysis.
By including the redshift information in a 2D field, the ARF observable keeps some information about the distribution of galaxies along the line of sight, which normally disappears when projecting the 3D galaxy density field on a 2D observable. As we have shown, ARF improve the usual 2D galaxy clustering analysis.

Another major interest of using angular observables is that they can easily be cross-correlated with other 2D observables. Indeed, the combination of 3D probes with 2D probes is not straightforward, especially when one has to properly take into account the covariances between them \citep[see e.g.][]{PassagliaManzotti2017,CameraFonseca2018}.
In this work, we used the CMB lensing field and its cross-correlation with our tomographic analysis of angular galaxy clustering and ARF.
We have shown that these cross-correlations improve the constraints, especially on the galaxy bias.
\citet{arfxkSZ_jonas} showed that the cross-correlation of the ARF field with the CMB temperature field can detect the kinematic Sunyaev-Zel'dovich (kSZ) effect at the $10\sigma$ level.

The point of this paper is not a detailed comparison between 2D and 3D clustering analyses, but rather an exploration of the added value of including ARF in cosmological studies of the large-scale structures, on top of the traditional angular galaxy clustering.
By their intrinsically different sensitivity to the cosmic density and velocity fields under the redshift shells, the ARF change the degeneracies between cosmological parameters, especially with respect to $\sigma_8$ and the galaxy bias, compared to the angular galaxy clustering.
This is due, as claimed in HMCMA, to the fact that angular galaxy clustering is sensitive to the first moment (the average) of matter density and velocity under the redshift shells, whereas ARF are sensitive to the variation of matter density and velocity along the line of sight inside these redshift shells.
Moreover, we have shown that the ARF and the angular galaxy clustering inside the same tomographic redshift bin are almost uncorrelated.
Due to this absence of correlation, by combining both we are able to break degeneracies and give tighter constraints on all the cosmological parameters we considered.

The results we obtained in our work can be considered as an optimistic setting for both galaxy and CMB surveys.
We restricted our analysis to the linear regime and we did not include any systematic effects that could impact our results and worsen the constraints.
\citet{arf_lett2} found that the impact of non-linear physics is more severe in angular galaxy clustering than in ARF. They found that a linear bias was sufficient to describe the ARF on scales larger than $60~h^{-1}$~Mpc, while it was not the case for angular galaxy clustering.
Indeed, ARF are built upon the average observed redshift along the line of sight in a redshift selection function.
This is intrinsically different to counting the number of galaxies in a given region in the universe, and consequently systematics and non-linearities affect each observable differently.
In future works, we plan to address systematics and non-linearities, aiming to model more realistic settings.
We expect that the impact of both systematics and non-linearities will depend on the survey and on the targeted galaxy sample, as ongoing work on existing galaxy surveys indicates.

We do not provide a detailed comparison with the forecasted constraints of the \Euclid survey published in \citet{EuclidCollaborationBlanchard2019}.
Indeed, our analysis considers a simplistic, linear model of the galaxy clustering.
In this context, our findings indicate that ARF brings significant cosmological information on top of the traditional angular galaxy clustering.
At best, our results with the angular galaxy clustering probe (\Dg) could be compared with the linear setting shown in Table~9 of \citet[][, first line]{EuclidCollaborationBlanchard2019}.
In that case, their probe is the 3D linear galaxy power spectrum, with a cutoff value at $k_{\rm max}=0.25~h$~Mpc$^{-1}$, in four different redshift bins. Their Fisher analysis accounts for more parameters describing the anisotropies in the power spectrum and the shot noise residuals.
This 3D probe is intrinsically different to the (2D) angular power spectrum tomography used in our work, in 20 Gaussian bins, which we limit to $k_{\rm max}=0.20~h~$Mpc$^{-1}$.
Our forecasts with \Dg for the errors on some parameters are tighter than theirs (by a factor of $\sim 2$ for $\sigma_8$), while for others we find the opposite situation (e.g. the reduced Hubble parameter $h$, whose uncertainty in \citealt{EuclidCollaborationBlanchard2019} is roughly one third of ours).

\section{Conclusion}
\label{sec:conclusion}

We show that the ARF are a promising cosmological observable for next generation spectroscopic surveys. We find that for our choice of binning the tomographic analysis of ARF retrieves more information than the tomographic analysis of the angular galaxy clustering.
We show that the joint analysis of both fields helps in breaking degeneracies between cosmological parameters, due to their lack of correlation and their different sensitivities to cosmology.
The improvement appears to be particularly significant for the \wcdm model. We show that the figure of merit for the $w_0$-$w_{\rm a}$ parameters was increased by a factor of more than ten when combining angular galaxy clustering with ARF.

Finally, we have seen that combining angular galaxy clustering with ARF provides tighter constraints on the galaxy bias parameters compared to the combination of angular galaxy clustering with CMB lensing. This shows that ARF are a very powerful probe of the distribution of matter, as they make it possible to break the degeneracy between $\sigma_8$ and the galaxy bias.
For future galaxy surveys, errors on the cosmological figure of merit will be dominated by systematic uncertainties and non-linearities, and ARF might provide a novel and complementary view on those issues.

In our analysis, we did not consider massive neutrinos. As the growth rate is particularly sensitive to them, we expect ARF to be a powerful tool to constrain the mass of neutrinos. We defer this detailed analysis to an upcoming work.

Simultaneously, from the LSS and CMB fronts, the coincidence in the acquisition of excellent-quality, extremely large data sets should enable the combination of standard analyses with new, alternative ones, like the one introduced in this paper. The combination of techniques and observables should work jointly on the efforts of identifying and mitigating systematics, and pushing our knowledge of cosmological physics to its limits.


\begin{acknowledgements}
The authors acknowledge useful discussions with
G.~Aricc\`o, G.~Hurier, and J.~Kuruvilla.
LL acknowledges financial support from CNES's funding of the Euclid project.
C.H.-M. acknowledges the support of the Spanish Ministry of Science and Innovation through project PGC2018-097585-B-C21.  NA acknowledges support from the European Research Council (ERC) under the European Union's Horizon 2020 research and innovation programme grant agreement ERC-2015-AdG 69556.
\end{acknowledgements}

\bibliographystyle{aa} 
\bibliography{biblio}

\appendix

\section{Correlation matrices}

\begin{figure}
    \centering
    \subfloat[$\Dg$]{\label{fig:param_corr_lcdm_desi_dg_full}\includegraphics[width=0.38\textwidth]{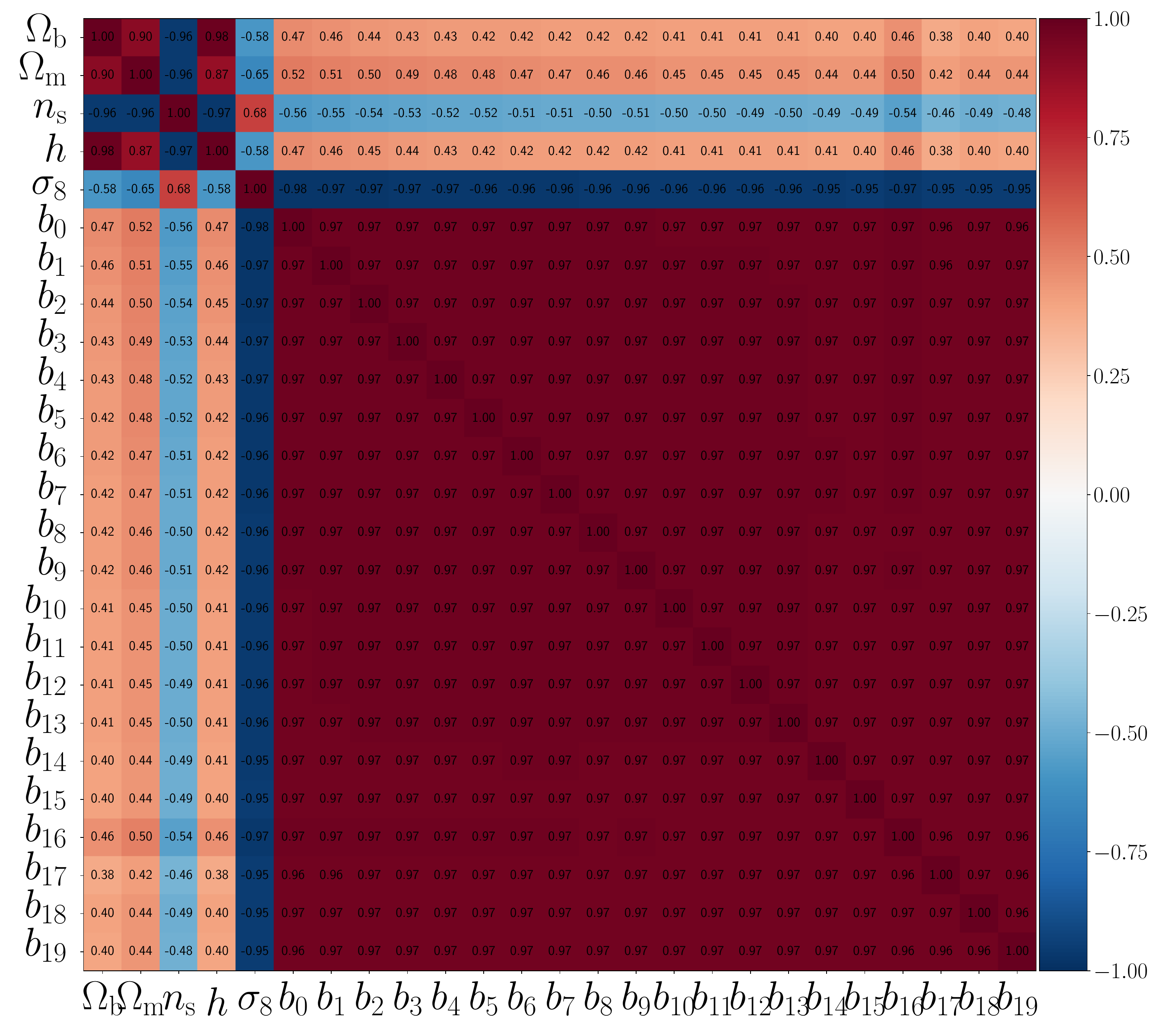}} \\
    \subfloat[$\Dz$]{\label{fig:param_corr_lcdm_desi_dz_full}\includegraphics[width=0.38\textwidth]{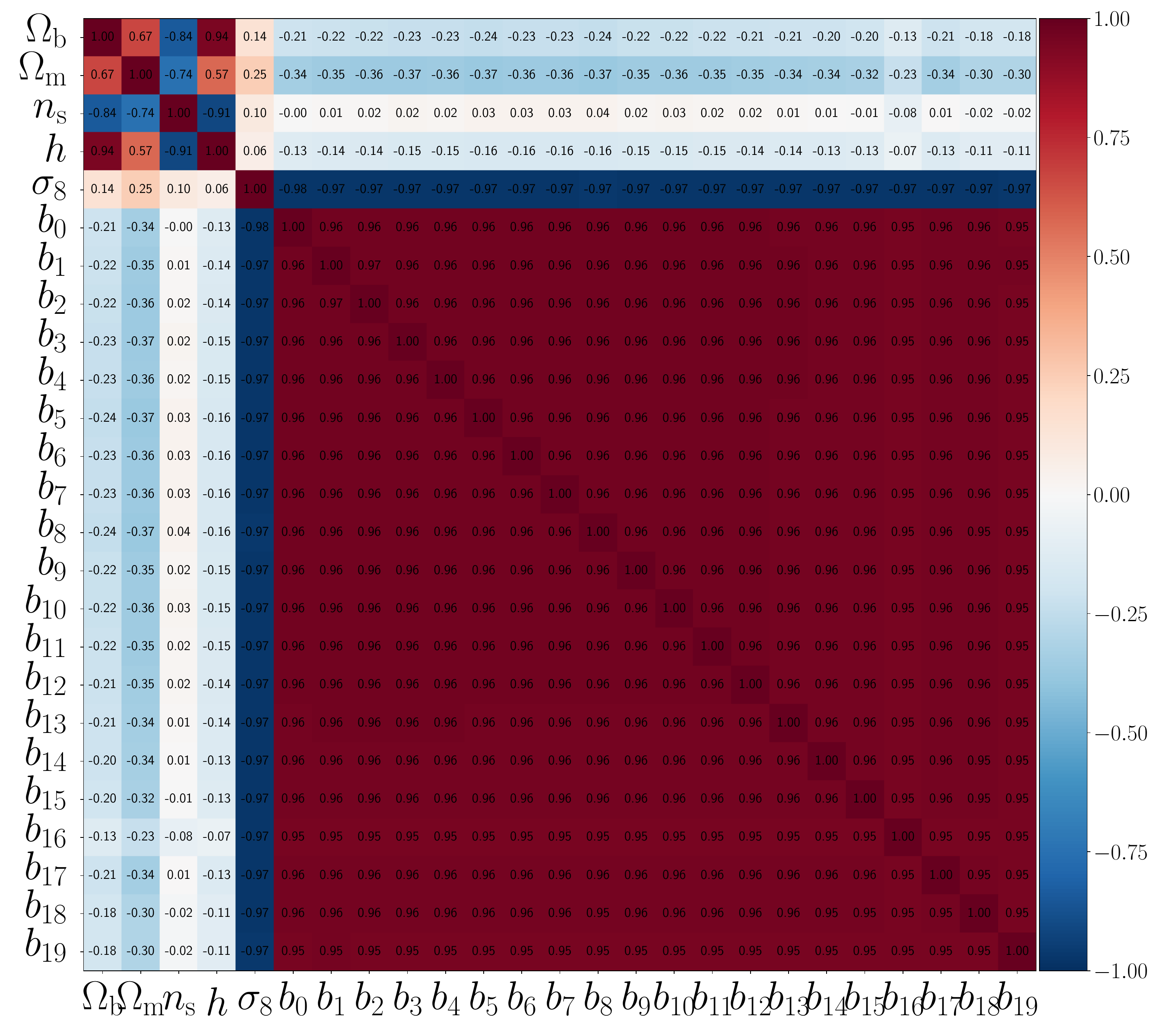}} \\
    \subfloat[$\Dgz$]{\label{fig:param_corr_lcdm_desi_dg-dz_full}\includegraphics[width=0.38\textwidth]{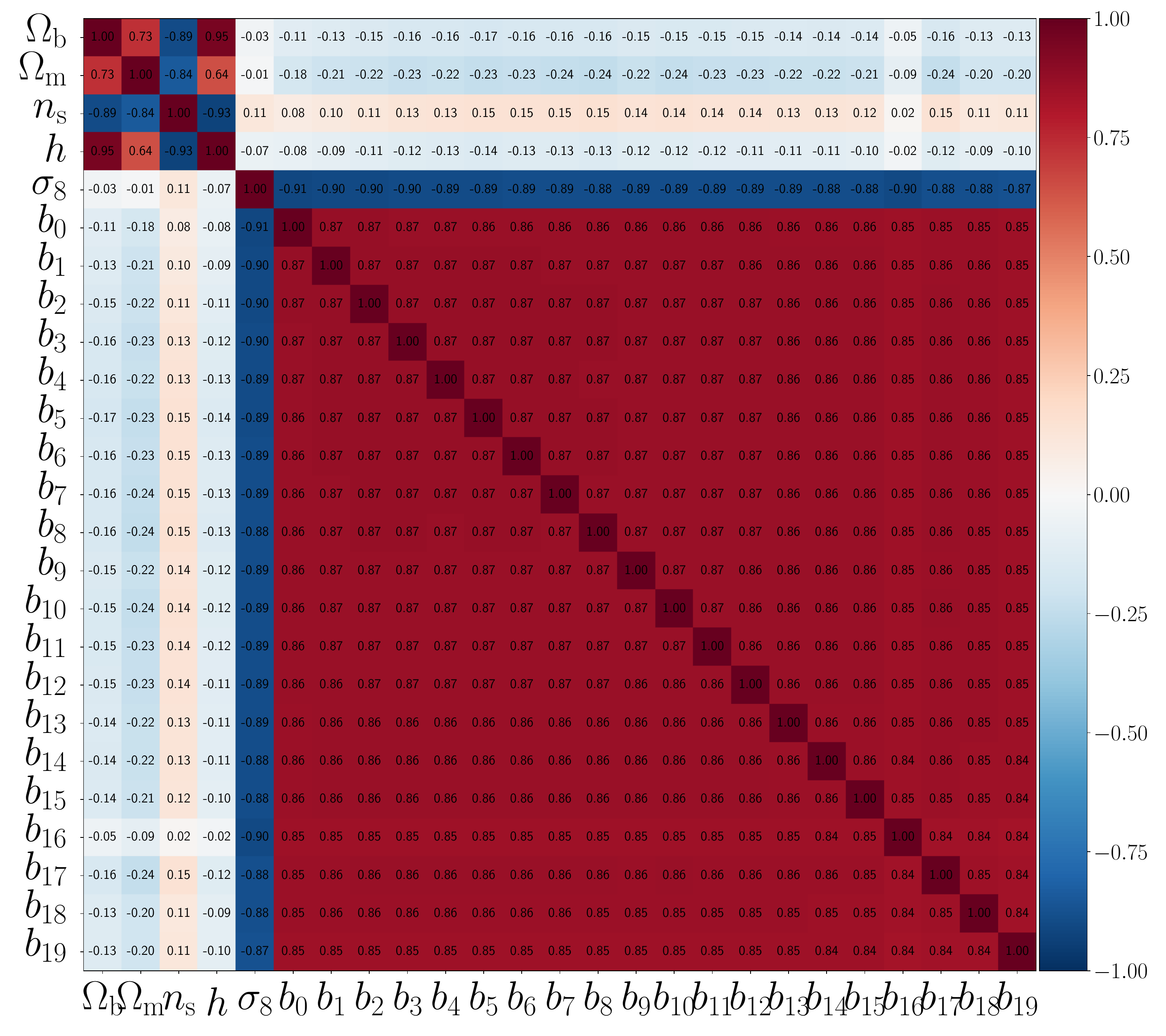}}
    \caption{Correlation between parameters of the \lcdm model for a DESI like survey. The top panel is for the angular galaxy clustering alone, the central panel is for ARF alone, and the bottom panel is for a combination of both observables. We see that the angular galaxy clustering and ARF have opposite correlation coefficients between cosmological parameters and the galaxy bias. The combination of both helps significantly to break degeneracies with the galaxy bias.}
    \label{fig:param_corr_lcdm_desi_full}
\end{figure}

The correlation matrices in Fig.~\ref{fig:param_corr_lcdm_desi_full} provide an alternative view of our results. It shows the correlation matrices for the five \lcdm parameters and the 20 galaxy bias parameters for a DESI-like survey.
We see the opposite correlation of the cosmological parameters $\Omega_{\rm b}$, $\Omega_{\rm m}$, $n_{\rm s,}$ and $h$ with $\sigma_8$ for angular galaxy clustering and ARF. This opposite correlation is mirrored in the correlations of those three cosmological parameters with galaxy bias parameters. This is expected, as $\sigma_8$ and bias are tightly correlated.
The different nature of the correlation of $\sigma_8$ and bias with the other cosmological parameters for angular galaxy clustering and ARF is critical for (partially)  breaking degeneracies when combining angular galaxy clustering with ARF.

\section{Results for the \wcdm model}

\begin{figure}
    \centering
    \subfloat[DESI]{\includegraphics[width=0.5\textwidth]{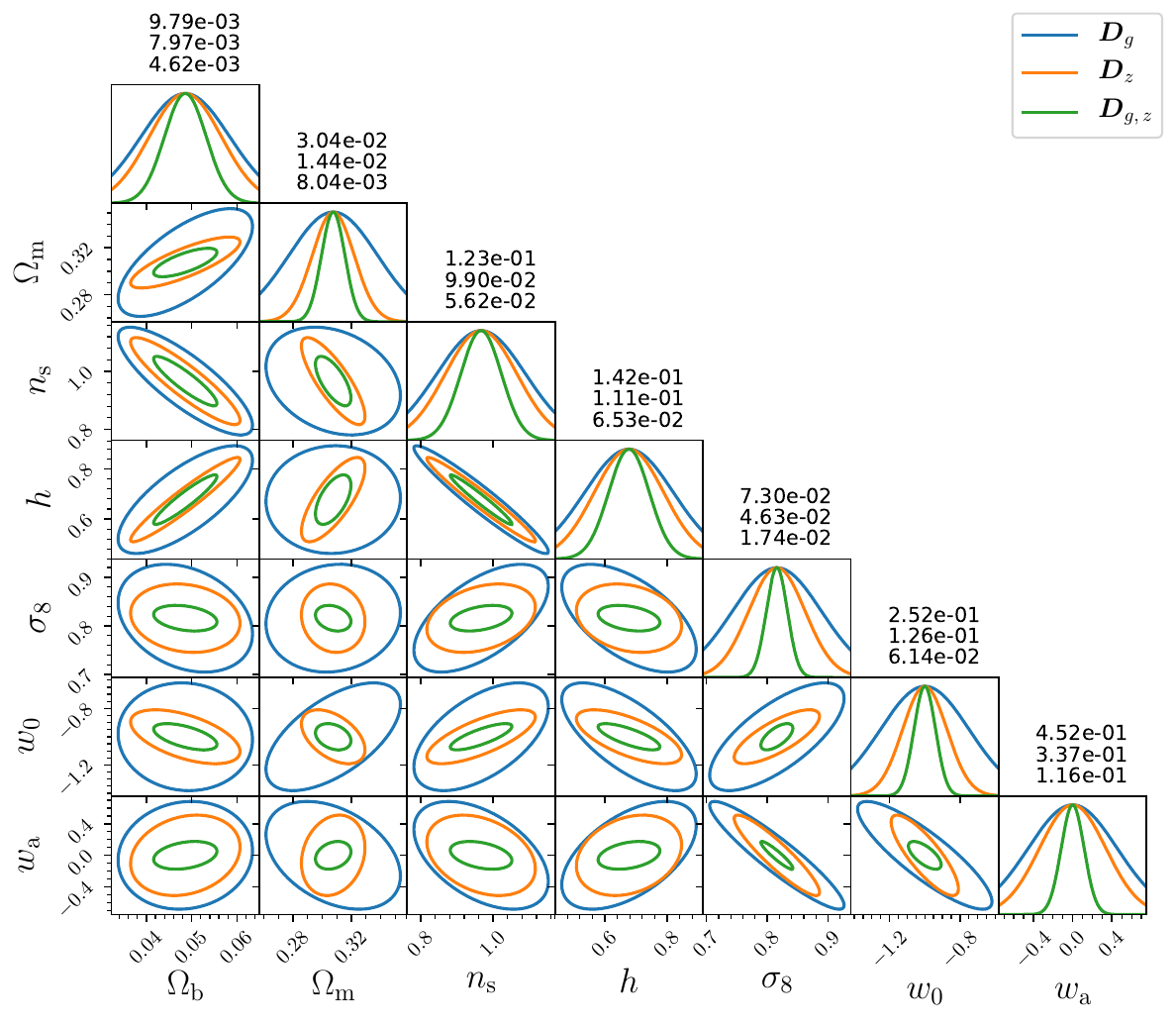}} \\
    \subfloat[\Euclid]{\includegraphics[width=0.5\textwidth]{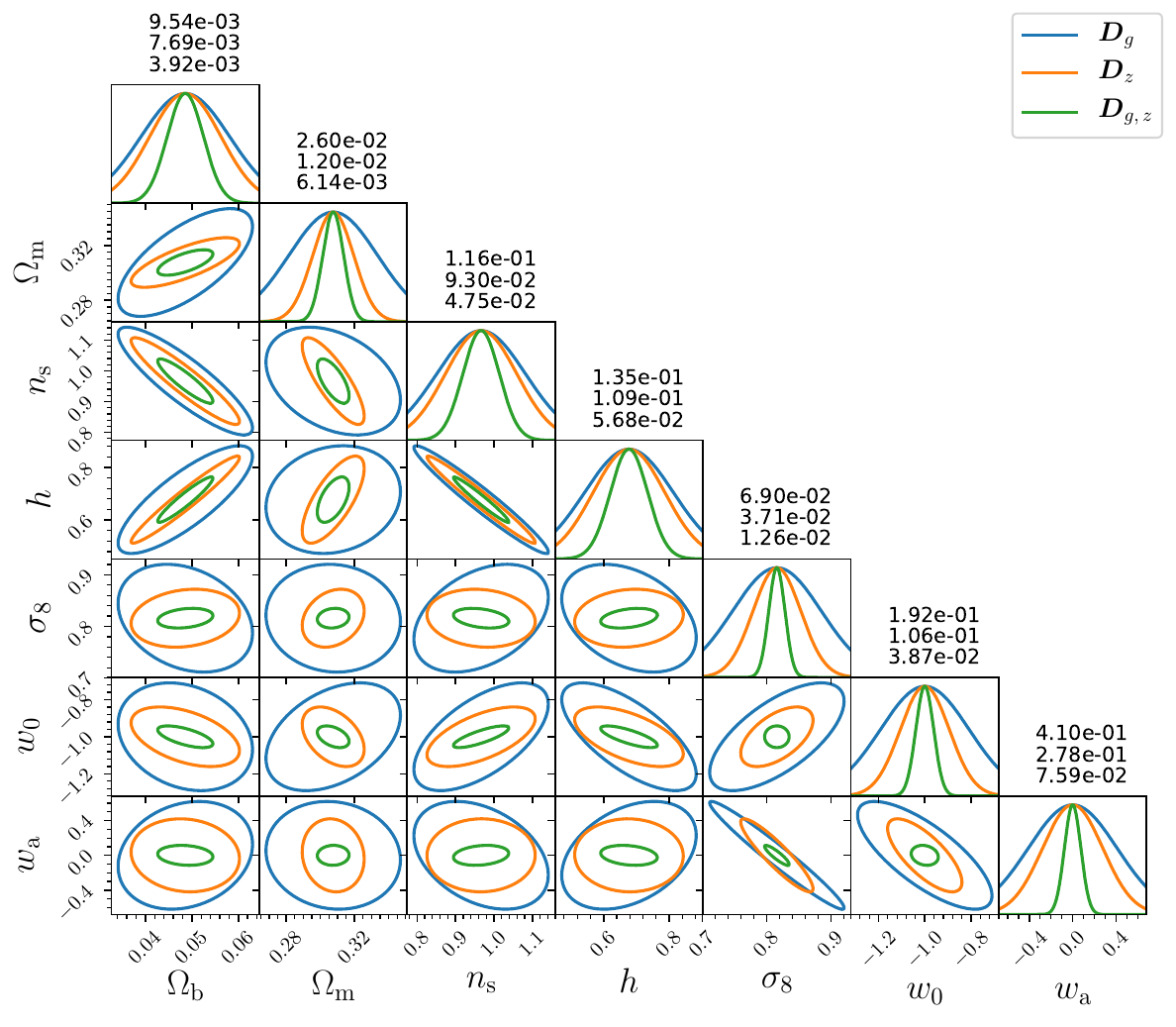}}
    \caption{Foreseen constraints ($1\sigma$ contours) in CPL cosmological extension \wcdm for a DESI-like survey (top) and an \Euclid-like survey (bottom), assuming 20 tomographic Gaussian bins of $\sigmaz=0.01$. The blue lines are the constraints for angular galaxy clustering alone (\Dg), the orange lines are for the ARF alone (\Dz), and the green line is a joint analysis of both fields (\Dgz). These contours are marginalised over the galaxy bias in the 20 redshift bins. The numbers above the parameter PDFs give the marginalised $1\sigma$ uncertainty of each parameter for each data vector.}
    \label{fig:ellipses_w0wa}
\end{figure}

We show the ellipses obtained with our Fisher analysis for the \wcdm model in Fig.~\ref{fig:ellipses_w0wa}, for a DESI-like and an \Euclid-like survey.
For many parameter pairs, the degeneracy direction (or ellipse orientation) for angular galaxy clustering and ARF are similar, although the resulting error ellipse in the joint \Dgz probe shrink very significantly in all cases.
As a result, foreseen uncertainties in the parameters are divided by a factor of at least two for all parameters.

\begin{figure}
    \centering
    \subfloat[\Dg]{\label{fig:param_corr_w0wa_desi_dg}\includegraphics[width=0.38\textwidth]{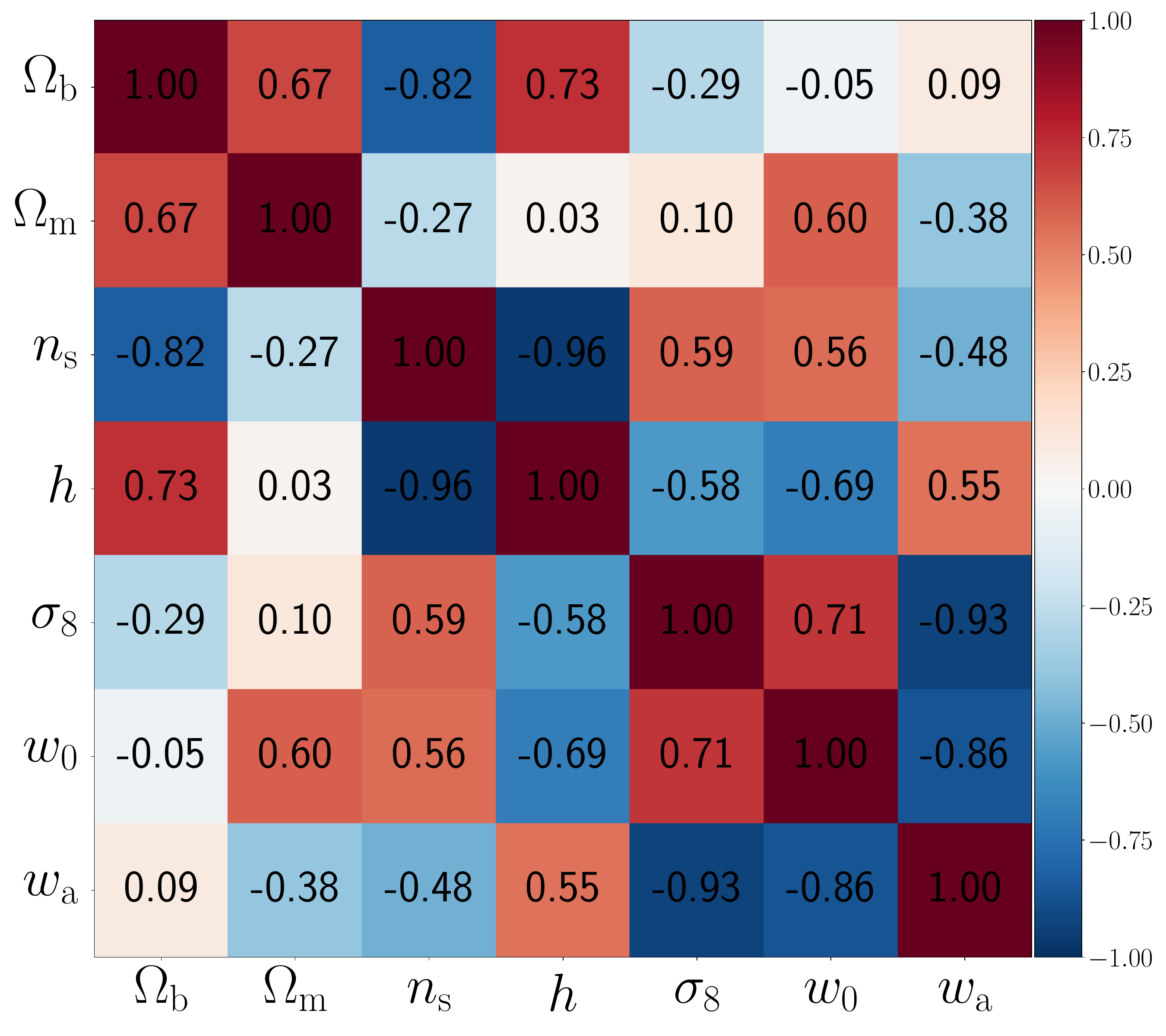}} \\
    \subfloat[\Dz]{\label{fig:param_corr_w0wa_desi_dz}\includegraphics[width=0.38\textwidth]{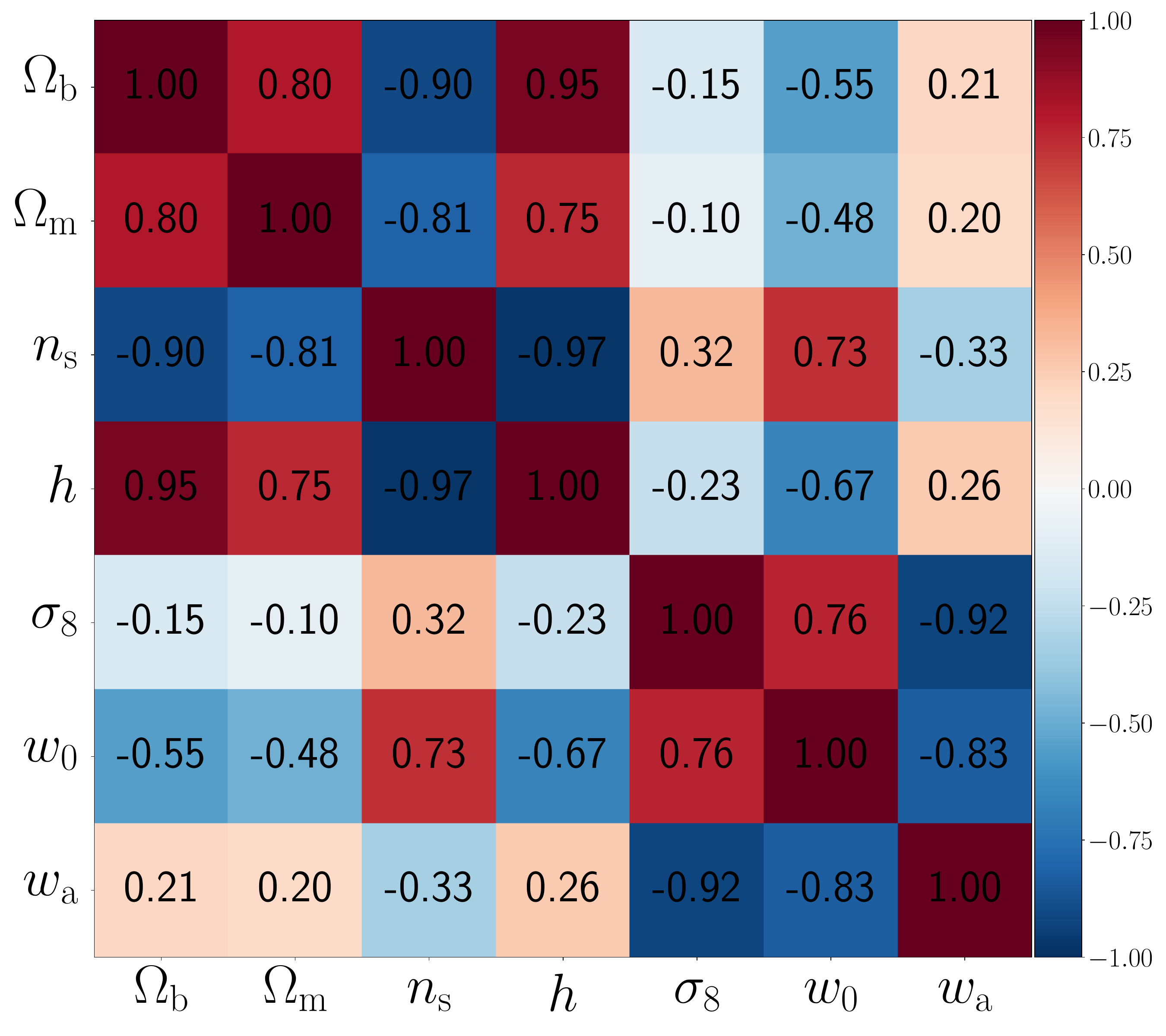}} \\
    \subfloat[\Dgz]{\label{fig:param_corr_w0wa_desi_dg-dz}\includegraphics[width=0.38\textwidth]{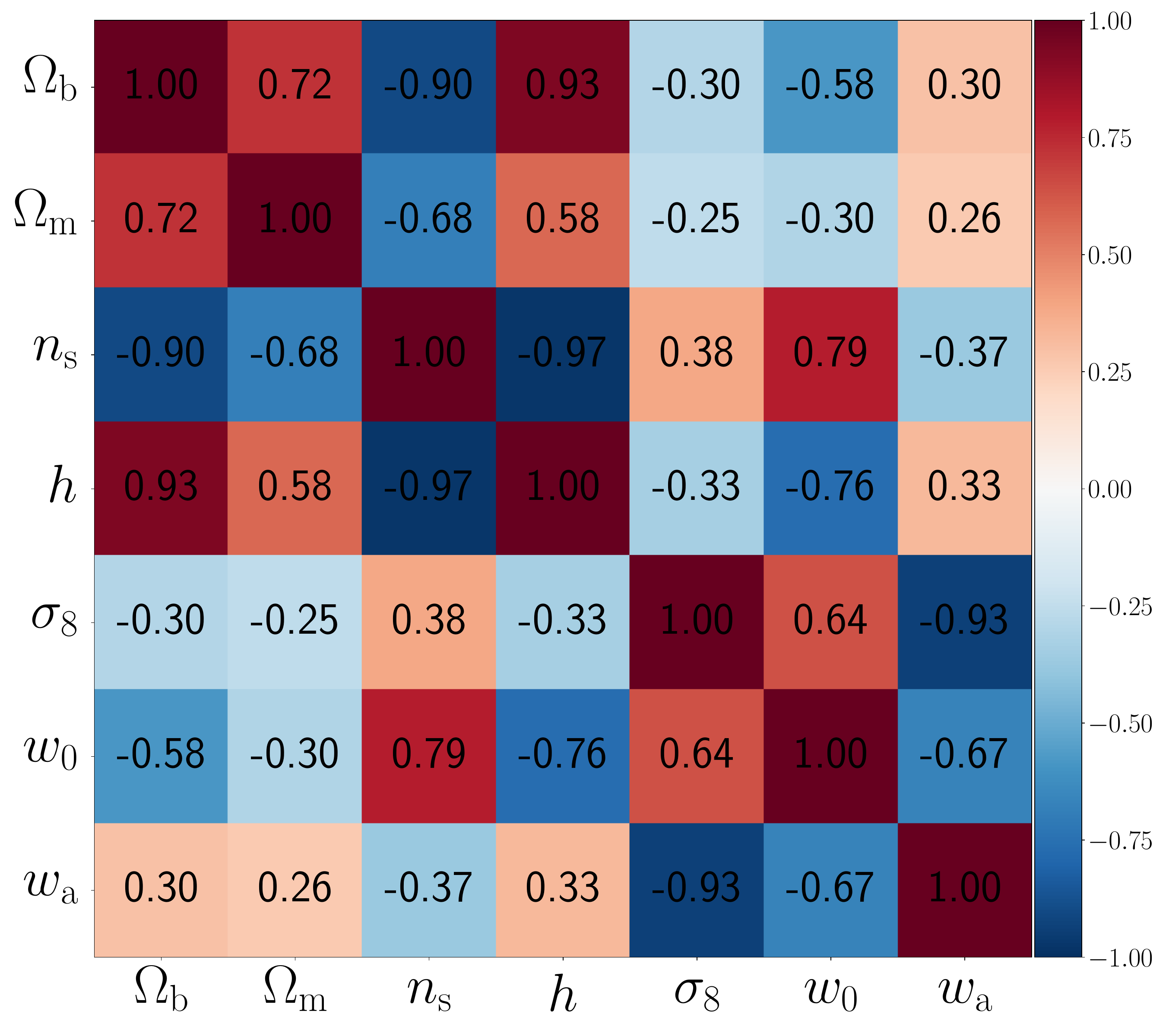}}
    \caption{Correlation between parameters of the \wcdm model, marginalised over the galaxy bias for a DESI-like survey. The top panel is for the angular galaxy clustering (\Dg), the central panel refers to ARF alone (\Dz), and the bottom panel is for a combination of both observables (\Dgz). We find that \Dz show different correlations compared to \Dg. The combination of both helps significantly to break degeneracies, as we can see in panel (c).}
    \label{fig:param_corr_w0wa_desi}
\end{figure}

In Fig.~\ref{fig:param_corr_w0wa_desi}, we show the correlation matrices for the \wcdm model.
It turns out that for $\Dz$ the new parameters, $w_0,\,w_{\rm a}$, together with $\sigma_8$, constitute an almost separate (or largely un-correlated) box with respect to all other parameters (see Fig.~\ref{fig:param_corr_w0wa_desi_dz}). This does not seem to be the situation for angular galaxy clustering $\Dg$ (Fig.~\ref{fig:param_corr_w0wa_desi_dg}), although this characteristic remains (to a great extent) for the joint observable set (\Dgz, Fig.~\ref{fig:param_corr_w0wa_desi_dg-dz}).

\end{document}